\documentstyle[aps,floats,psfig,multicol,pre,amstex]{revtex}

\newcommand{\at}{{\char '100}}

\newcommand{\DEF}{\stackrel{\mathrm{def}}{=}}
\newcommand{\imat}{{\mathrm{i}}}    
\newcommand{\hbare}{\hbar_{\mathrm{eff}}}  

\newcommand{\opp}{\hat{p}}
\newcommand{\opq}{\hat{q}}

\begin{document}

\draft

\title{Chaos assisted tunnelling with cold atoms}
\author{
       A. Mouchet$^1$\thanks{mouchet\at celfi.phys.univ-tours.fr},
   C. Miniatura, R. Kaiser$^2$\thanks{miniat,kaiser\at inln.cnrs.fr},
     B. Gr\'emaud, D. Delande$^3$\thanks{gremaud,delande\at
spectro.jussieu.fr}
       }
\address{        $^1$Laboratoire de Math\'ematiques 
                            et de Physique Th\'eorique 
                            \textsc{(cnrs upres-a 6083)}.
                            Ave\-nue Monge, Parc de Grandmont 37200
Tours, 
                            France.\\
         $^2$Institut Non Linéaire de Nice \textsc{(cnrs umr 7618)},
                                1361, route des Lucioles,
                                Sophia Antipolis,
                                F-06560 Valbonne, France.\\
 $^3$Laboratoire Kastler-Brossel \textsc{(cnrs umr 8552)}, Universit\'e
                               Pierre et Marie Curie, 4, place Jussieu,
                               F-75005 Paris,  France. 
        }

\date{\today}

\maketitle

\begin{abstract}
In the context of quantum chaos, both theory and numerical analysis
 predict large
 fluctuations of the tunnelling transition
probabilities when irregular dynamics is present at the classical level.
We consider here the non-dissipative quantum evolution of 
cold atoms trapped in a time-dependent modulated periodic potential 
generated by 
two laser beams.  We
give some precise guidelines for the observation of 
 chaos assisted tunnelling between invariant phase space structures 
 paired
by time-reversal symmetry.

\end{abstract}


\bigskip
                     
\section{Introduction}
 
During the seventies and the eighties, it became gradually clear that
classical Hamiltonian chaos profoundly affects 
the  temporal evolution and the spectral  properties 
of the corresponding quantum system as 
compared to the integrable case~\cite{Giannoni+91a}.
Some of these features 
(dynamical localization, 
scars of periodic orbits~\cite{Heller84a}, etc) 
share striking similarities with concepts
 originating from condensed matter such as weak and strong 
localization~\cite{Akkermans+95a}. In fact these
phenomena can be recast in terms of wave transport in disordered
media, (quasi)-randomness being of statistical or dynamical origin.
In this  context, it is important to 
understand the mechanisms 
underlying a key feature of wave propagation which has no
classical analog: tunnelling.

Tunnelling refers to any wave  process which is classically
forbidden to \emph{real} solutions of Hamilton equations. For one-dimensional
(1D)
 autonomous systems, it is well known that the quantum 
tunnelling probability 
through an
energetic barrier can be
evaluated semiclassically with the help of classical 
\emph{complex} solutions 
of Hamilton 
equations~\cite{Messiah65a,Balian/Bloch74a}. 
The direct generalization of this procedure to higher
dimensional  systems is straightforward for separable dynamics 
but is already subtle for  still integrable,
 but no longer separable,
dynamics~\cite{Wilkinson86b,Wilkinson/Hannay87a}. 
In the generic case 
of chaotic dynamics,  it even proves 
 extremely hard to
handle and the situation, until recently, seemed hopeless.
This is so because, in the presence of chaos, the analytical and
topological 
properties of the classical
\emph{complexified}  phase space are far from trivial.
  During the last ten years however, theoretical and numerical 
investigations on autonomous 2D and time-dependent 1D Hamiltonians
systems have 
started to highlight
some
mechanisms~\cite{Wilkinson86b,Wilkinson/Hannay87a,Lin/Ballentine90a,Bohigas+93a,Bohigas+93b,Tomsovic/Ullmo94a,Creagh/Whelan96a} and 
much insight
has been gained on the influence of such classical
 non-separable dynamics. Experimental
 evidence of such mechanisms, which is still lacking,
 would be of great interest especially in the light 
of the subtle interplay between interferences and disorder.

 In this paper, we consider 1D time-dependent 
 dynamics, one of the simplest case where
irregular motion can appear, and we study  
chaos assisted tunnelling.  Our effective Hamiltonian 
model, which is derived from an experimentally
achievable situation, exhibits three 
main properties. First, 
its classical dynamics is time-reversal invariant.
Second it is
controlled by a single real external parameter $\gamma$ (for $\gamma=0$
 the dynamics is integrable and chaos develops more and more in
phase space as $\gamma$ is increased). Third, there exists 
in phase space, for a whole  continuous range of~$\gamma$,
 a pair of stable 
islands~$\mathcal{I}_{+}$ and~$\mathcal{I}_{-}$ which are time
reversed images of each other. By stable islands we mean the set 
of regular classical 
trajectories in phase space which stay near 
 a stable equilibrium point or near a stable periodic orbit of the
system.
In this case,
no real classical orbit started in one of these islands can
 go into the other one. However, the quantum
dynamics of a wave-packet, initially prepared in one island, will
display a periodic behavior. The wave-packet jumps from one island to
its time reversed image~\cite{Averbukh+95a}.
  In the quantum spectrum
this tunnelling process appears via the 
existence of non
degenerate energy doublets whose splitting give the inverse of the
tunnelling time  between~$\mathcal{I}_{+}$ 
and~$\mathcal{I}_{-}$. Varying~$\gamma$ slowly modifies 
the geometry of the islands themselves.
 The crucial point is that it will drastically change the classical
dynamics
for some initial conditions lying between the islands. For~$\gamma$
small 
enough, the chaotic layers are too small to play a significant role
at~$\hbar$ scales and hence cannot influence the quantum behaviour of
the 
system
which is essentially still regular.
For larger values,  but still
 before the stable islands are completely destroyed, there is a chaotic
regime where varying~$\gamma$ or $\hbar$ 
(by $\hbar$ we mean here Planck's constant 
divided by some typical classical action)
alone induces large fluctuations, on several orders of magnitude,
of the doublet splittings around their mean value. This in turn 
means large fluctuations of the
tunnelling periods. These wild fluctuations
induced by small changes of any  parameter are a
signature of the so-called ``chaos assisted tunnelling'' regime.
It has been
extensively studied both theoretically and 
numerically in the situation described 
above~\cite{Lin/Ballentine90a,Bohigas+93a,Bohigas+93b,Tomsovic/Ullmo94a} 
but has not been yet observed in 
real experiments, the main reason being its extreme sensitivity 
to small changes in the
classical dynamics.
 Any \emph{uncontrolled} variation of $\gamma$, be it
 noise or dissipation, will dramatically swamp or destroy the signal.
The
observation of this highly fluctuating
tunnelling regime thus requires both an
accurate control of the dynamics, of the preparation 
 of the initial state and of the
analysis of the final state.

Atom cooling techniques~\cite{Arimondo+92a}
 provide systems which fulfill all these requirements. 
 They allow an accurate manipulation and
control of internal and external degrees of freedom and are
now a useful tool to produce situations where the wave character
of the atomic motion is 
essential~\cite{Berman96a}. 
A great variety of potentials can be produced to influence the
atomic motion, be it by means of inhomogeneous magnetic fields, material
gratings or laser light. Optical
 lattices with crystalline or 
quasi-crystalline
order~\cite{Weidemuller+95a,Guidoni+97a,Drese/Holthaus97a} 
can be easily
produced where atoms mimic 
situations usually
encountered in condensed 
matter~\cite{BenDahan+96a,Niu+95a}.
 Dissipation (spontaneous emission and atom-atom interaction) 
is easily controlled
 and coherence times of the order of 10 ms
can be achieved. This is why cold atoms are a unique tool to
study transport properties of waves, be it quantum
chaos~\cite{Moore+94a} 
or weak localization~\cite{Labeyrie+99a,Jonckheere+00a}.

The general organization of this paper is the following. 
In section~\ref{sec:experimentalsetup}  we
explain the origin of the effective Hamiltonian
 for the experimental situation
 under consideration. In section~\ref{sec:classicaldynamics} 
we study the corresponding classical dynamics 
and show why this effective Hamiltonian is relevant for chaos assisted
 tunnelling. In 
section~\ref{sec:quantumdynamics} we quickly review some of the
usual theoretical techniques when dealing with both space and time 
periodic quantum dynamics. We also
illustrate  how some quantum spectral properties 
 have a natural classical interpretation.
In section~\ref{sec:cat} we show, with the help of numerical 
experiments,
 how chaos assisted tunnelling
arises in our system and then explain how to observe it 
in a real experiment.
Section~\ref{sec:conclusion} is devoted to some concluding remarks.

\section{Effective Hamiltonian}
\label{sec:experimentalsetup}
\subsection{Light shifts}

The very basic physical mechanism underlying 
our forthcoming discussion 
is the following: when an atom is exposed to monochromatic light, 
its energy levels are shifted by the interaction. 
 These level shifts originate from the polarization 
energy of the atom in the incident light field and are called the
light shifts~\cite{CohenTannoudji+88a}. In the dipolar 
approximation, they only depend on the field intensity
value at the center-of-mass position of the atom. 
If the field intensity is space-time 
dependent, then a moving atom will experience dipolar forces: 
inhomogeneous light shifts act as potentials and alter the 
center-of-mass 
motion of the atom. By 
appropriately tailoring the space-time dependence of the 
light field, one can then produce a great variety of potentials 
for the external atomic motion as evidenced by the atom cooling 
industry. Note, however, that the atom-light interaction is also
responsible for a dissipative phenomenon (real absorption of a
photon followed by spontaneous emission) which shortens the
temporal coherence of the atomic wave function. By using a laser light
far detuned from any atomic resonance, it is possible to control
this stray phenomenon and maintain it at a reasonably low rate.

In the following, we shall describe a simple physical 
situation for atoms where chaos 
assisted tunnelling should show up.

\subsection{Experimental configuration}

\begin{figure}[!ht]
\center
\psfig{file=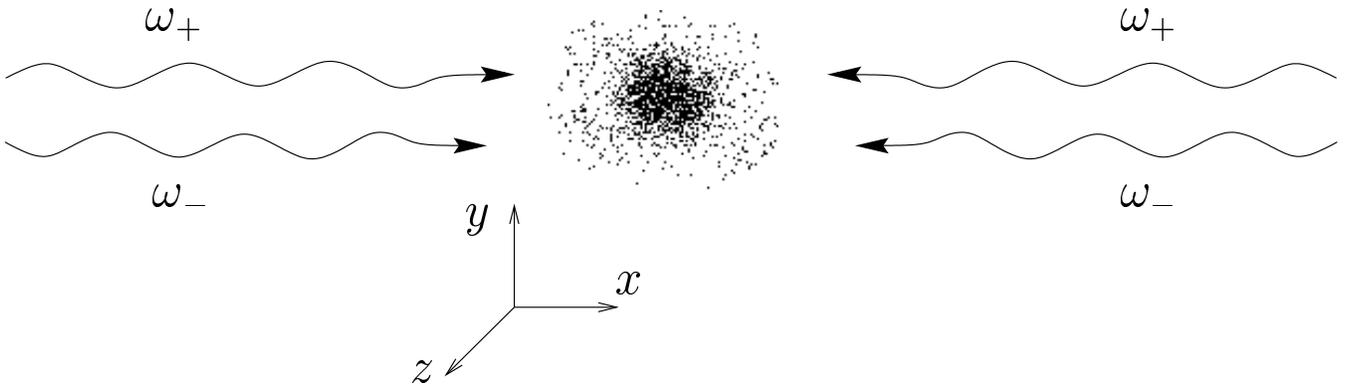,width=18cm}
\caption{\label{fig:piegeos}Experimental configuration under
consideration:
a cloud of two-level atoms is exposed to two monochromatic standing waves with
frequencies
$\omega_\pm = \omega_L \pm \delta\omega/2$ $(\delta\omega\ll\omega_L)$.
 All fields are
linearly polarised along the same direction and are sufficiently
far-detuned
from the
atomic resonance so that dissipation effects can be ignored.}
\end{figure}

Although the internal structure (hyperfine Zeeman sub-levels) are
of major importance in the atom cooling techniques, we will here for
simplicity model the atom by a two-level system (as indeed, only one
optical transition usually governs the dynamics). 
We consider a dilute sample of identical (but independent) 
two-level atoms propagating in the light field configuration
 created by two monochromatic standing waves 
with frequencies $\omega_\pm=\omega_L\pm\delta\omega/2$ 
where~$\delta\omega\ll\omega_L$.
 We denote
 by $|g\rangle$ and $|e\rangle$ the ground-state and 
the excited state of each atom, these levels being connected by 
an electric dipole transition of angular
 frequency $\omega_{\mathrm{at}}$ and width~$\Gamma$. 
  All atoms are supposed to be initially 
prepared in their ground state. 
Each standing-wave is produced along the $x$ axis by two
counter-propagating 
laser beams and we suppose all fields
to be linearly polarized along the $z$ axis (see
figure~\ref{fig:piegeos}). 
After a suitable choice of space-time origin, the total electrical 
field strength is
\begin{equation}
        E(x,t)=\big[E_+\cos(\omega_+t)+E_- \cos(\omega_-t)\big] \cos(k_Lx)
\end{equation}
where $E_\pm$ are the field strengths of the two standing-waves. 
At this point we have neglected the difference in wave-vectors 
of the standing-waves. 
For this to hold, it is sufficient to assume that
the atomic sample size is small enough. Typically, the difference in the
$k$ vectors will be of the order of $10^{-9}$ or less (see below),
so that this requires the atomic cloud to be smaller than typically
few kilometers, which is amply satisfied in a standard magneto-optical trap.

\subsection{Dimensionless effective Hamiltonian}\label{subsec:effham}

The effective Hamiltonian which describes the atomic motion is 
derived in appendix~\ref{app:effham} under some common and
well-controlled 
approximations. It acts in the Hilbert space of a one-dimensional system
which is simply the $x$ component (position) of the center
of mass of the atom
(which means that the internal degree of freedom as well as the
$y$ and $z$ coordinates can be eliminated, see Appendix). It reads: 
\begin{equation}\label{eq:ham_eff}
        H=\frac{p_x^2}{2M}-V_0\cos(2k_Lx)[\theta+\cos(\delta\omega\, t)]
\end{equation}
where $V_0\DEF-\hbar\Omega_+
\Omega_-/8\delta_L$ and 
$\theta\DEF(\Omega_+/2\Omega_-)
+(\Omega_-/2\Omega_+)$, with $\delta_L=\omega_L-\omega_{at}$ the
detuning with respect to the atomic frequency 
and $\Omega_{\pm}=dE_{\pm}/\hbar$ ($d$ being the atomic dipole strength).
Without loss of generality we will assume~$V_0$ to be positive since, if~$V_0$
is negative, it is sufficient to shift~$x$ by~$\pi/2k_L$ to recover this case.

In the following, it will prove convenient to work with dimensionless
quantities. Rescaling quantities through 
$\tau\DEF\delta\omega\, t$, $q\DEF2k_Lx$, 
$p\DEF(2k_L/M\delta\omega)p_x$, 
$\gamma\DEF(4k_L^2/M\delta\omega^2)V_0$ 
and $H_{\mathrm{eff}}\DEF(4k_L^2/M\delta\omega^2)H$, 
then yields the dimensionless effective Hamiltonian:
\begin{equation}\label{eq:ham_red}
H_{\mathrm{eff}}
        =\frac{p^2}{2}
        -\gamma \ (\theta + \cos\tau )\ \cos q\;.
\end{equation}
Such an Hamiltonian describes the dynamics of a periodically 
driven pendulum. The associated quantum canonical commutation 
relation is 
$[q,p]=i\hbar_{\mathrm{eff}}$ and we get 
$\hbar_{\mathrm{eff}}=8\omega_R/\delta\omega$ 
where $\omega_R=\hbar k_L^2/2 M$ 
is the atomic recoil frequency
 and  $\delta\omega$ is the beating frequency between the laser 
waves.

Such an effective Hamiltonian clearly exhibits two of the three 
properties mentioned in the introduction: the corresponding 
classical dynamics 
is governed by a single classical parameter, the dimensionless 
coupling strength 
$\gamma$, and is invariant 
under 
 time-reversal symmetry 
$(p,q,\tau)\mapsto (-p,q,-\tau)$. 
It is worth mentioning that the semiclassical limit 
$\hbar_{\mathrm{eff}} \to 0$ is realized here by increasing the beating
 frequency $\delta\omega$ 
between the two laser waves.
 
With our field configuration, only $\theta\ge 1$ can be achieved. 
As a slight generalization, we extend the range of 
$\theta$ to any positive value since one can
 design other field configurations where $\theta\le 1$ occurs. For
example, 
$\theta=0$ yields the Hamiltonian
studied in~\cite{Averbukh+95a} in a different 
context.

\subsection{Orders of magnitude}

Let us give some typical experimental parameters.
 For Rubidium atoms, the atomic parameters are $M=85\,\mathrm{amu}$, 
$\lambda_{\mathrm{at}}=2\pi c/\omega_{\mathrm{at}}=0.78\,\mathrm{\mu
m}$, 
$\Gamma/2\pi = 6\,\mathrm{MHz}$, 
$\omega_R/2\pi=3.8\,\mathrm{kHz}$ and saturation intensity 
$I_{\mathrm{sat}}=1.6\,\mathrm{mW/cm^2}$. 
Using far-detuned laser beams ($2\delta_L/\Gamma = 10^4$) 
 focussed down to $500\,\mathrm{\mu m}$ (power $100\,\mathrm{mW}$), 
with a beating 
frequency $\delta\omega/2\pi = 60\,\mathrm{kHz}$, lead to 
$\gamma = 0.4$ and $\hbar_{\mathrm{eff}} =0.05$. With such 
values, spontaneous emission can be neglected up to times of 
the order of few $\mathrm{ms}$. It is worth noticing the tiny energies
which 
come into play ($V_0\sim 5\,\mathrm{neV}$), by several orders of
magnitude
smaller than the typical ones for mesoscopic systems.

\section{Classical dynamics}\label{sec:classicaldynamics}
\subsection{Poincar\'e surface of section}
A Poincar\'e surface of section provides the usual tool for visualizing
 the classical dynamics~\cite{Lichtenberg/Lieberman83a}. 
As $H_{\mathrm{eff}}$ is $2\pi$-periodic both in
time and space, this surface of section simply consists 
in the whole phase space itself 
(which has the topology of a cylinder) where trajectories 
$(p(\tau),q(\tau))$ 
are seen stroboscopically at every time period $2\pi$. 
In the following, without any substantial loss of generality, 
we will restrict our analysis to the
 case~$\theta=1$ which is easily
experimentally achieved when the standing waves have the same 
field strengths.

\begin{figure}[!ht]
\center
\psfig{file=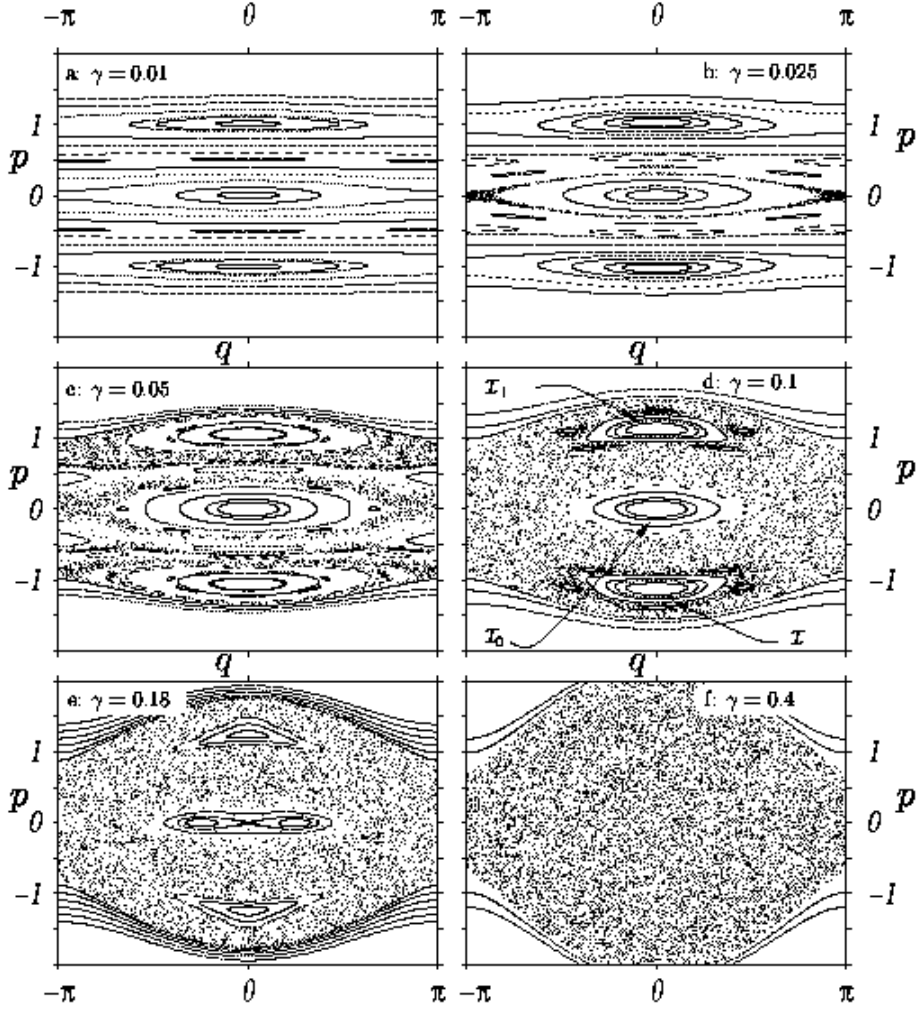,height=19cm}
\caption{\label{fig:3yeux} Stroboscopic plots of trajectories in phase
 space for different initial conditions at~$\tau=0$ and
different~$\gamma$'s. 
The classical dynamics is governed by Hamiltonian~\eqref{eq:ham_red}
with 
$\theta=1$. At low $\gamma$ values, resonance islands are visible
separated by quasi-free motion. As $\gamma$ increases, the resonance
islands
grow and chaos appears close to the separatrices. The situation of
interest
for chaos assisted tunnelling is when two symmetric islands are separated
by
a chaotic sea, as $\mathcal{I}_+$ and ${\mathcal{I}}_-$ in (d) and (e).
}
\end{figure}

Figure~\ref{fig:3yeux} shows stroboscopic plots
 of phase space orbits for different
 $\gamma$'s. 
For $\gamma=0,$ $p$ is a constant of motion, so that the
system is integrable and the surface of section is composed
of horizontal lines.
For $\gamma$ weak enough (fig.~\ref{fig:3yeux}-a), the orbits remain
confined 
 to invariant curves. These invariant curves stratify the whole 
 phase space and the dynamics appears regular. One can clearly see 
 well separated stability islands, each being bordered by a separatrix. 
 This is the situation encapsulated in the \textsc{kam} theorem 
 for near-integrable 
 motion: although no globally-defined constants of motion exist, 
 some invariant curves
 can still be 
 constructed which order the dynamics. As $\gamma$ is increased
 (fig.~\ref{fig:3yeux}-b), 
 some of more and more invariant curves are broken and  chaotic
 layers start to 
 spread around  separatrices. These layers fill some portion of phase 
 space but motion is still predominantly confined to invariant curves. 
 Above some coupling threshold (fig.~\ref{fig:3yeux}-c 
to~\ref{fig:3yeux}-e), stochastic orbits invade 
 phase space and the surviving stability islands are surrounded by a 
 connected chaotic sea. This occurs for $\gamma\sim0.1$. The phase space 
 structure in this regime is typical of a mixed dynamics where regular 
 orbits co-exist with stochastic ones. If $\gamma$ is 
 increased further (fig.~\ref{fig:3yeux}-f), the stability 
islands disappear (or are too 
 small to be seen at this scale)
 and one gets global chaos. We note 
 however that, even in this situation, the chaotic portion of phase
space 
 is still bounded by invariant curves which means that chaos can only
fully 
 develop within some range of momentum $p$.

\subsection{Resonances}\label{subsec:resonances}

At this stage, let us rewrite the effective 
Hamiltonian as follows:

\begin{equation}\label{eq:ham_red2}
        H_{\mathrm{eff}}=H_0+\gamma H_1
        =\frac{p^2}{2}\,-\gamma\cos q
         -\frac{\gamma}{2}\cos(q+\tau)
         -\frac{\gamma}{2}\cos(q-\tau)
\end{equation}

The physical interpretation of the various terms is rather simple:
the two counter-propagating laser beams at frequency $\omega_+$
create a stationary wave which, in turn, creates for the atom
an effective optical potential proportional to the
square of the modulus of the electric field in the standing wave,
hence the $\cos q$ dependence (it is actually rather
$1+\cos q$, but the constant term does not play any role in
the dynamics). The same effective potential is due to the
standing wave created by the two $\omega_-$ counter-propagating beams.
A pair of counter-propagating beams at frequencies $\omega_+$ and
$\omega_-$ does not create a standing wave in the lab frame. However,
in a frame sliding at constant velocity
$v_0=(\omega_+-\omega_-)c/2\omega_L,$
(=1 in rescaled units)
the two laser beams are shifted in frequency by Doppler effect
and appear to have equal frequency, building another stationary
wave and yet another effective optical potential. In the lab frame,
this appears as a modulated optical potential moving at velocity $v_0.$ 
By symmetry, there are two such effective potentials moving
either to the right or to the left. These are the $\cos(q\pm\tau)$
terms in the Hamiltonian.

This form of the Hamiltonian is in 
order to point out the perturbative terms which may be resonant
with the unperturbed frequencies. When~$\gamma=0$, the system is 
integrable since we recover free motion: $H_{\mathrm{eff}}$ reduces 
to $H_0=p^2/2$ and $(p,q)$~are exact action-angle variables. 
For $\gamma>0$, the absence of any 
constant of motion
generates chaos. Stroboscopic plots of phase-space trajectories 
 are no more constrained to follow lines of
constant~$H_0$ but generically fill densely a two dimensional volume 
in phase space. As long as $\gamma$ is small enough, these volumes
remain
thin enough not to be distinguished from regular lines at the scale
of finite precision of the measurements and/or the calculations 
(\textit{cf.}~figure~\ref{fig:3yeux}-a).
 Nevertheless, for higher values of~$\gamma$, some chaotic layers can be
seen
(\textit{cf.}~figure~\ref{fig:3yeux}-b)
between regular regions. They consist of portions of phase space where 
trajectories are exponentially sensitive on initial conditions. 
From classical 
first-order perturbation
 theory~\cite[chap.~2]{Lichtenberg/Lieberman83a}, 
we can infer that a term of the form $A \cos(sq-r\tau)$,
where $(s,r)$ are integers, will create a resonance of width 
$\Delta p = 4\sqrt{A}$ around the point $p=r/s$. In our case, $s=1$
and 
there exist 
only three such resonances. They are located at $p=0$ 
($r=0$) and at 
$p=\pm1$ ($r=\pm1$). This can be seen in 
figure~\ref{fig:3yeux} (a to e). 
For each resonance there exists one stable and one unstable periodic 
orbit with period approximately $2\pi |r|.$ In the stroboscopic plot
of the surface of section, they appear as a stable and unstable fixed
points and give rise locally to the well-known phase-space portrait 
of a pendulum. In the following, we will 
denote by~${\mathcal{I}}_0,{\mathcal{I}}_+,{\mathcal{I}}_-$
the three stable islands associated with $r=0,+1,-1$ respectively. 
The physical interpretation of these three resonances is simple:
each resonance is associated with one of the modulated potentials
(either static or moving) described above. For example,
the fixed point at the center of the ${\mathcal{I}}_+$ resonance
is associated with a periodic orbit where the atom moves at almost
constant velocity $v_0,$ being is fact trapped in the minimum of the
moving optical potential. The other two components of the
potential appear along this orbit as rapidly varying potentials
which are adiabatically averaged to constant values.
As the atom can be trapped in any of the 3 modulated potentials, we
obtain three stable periodic orbits at the centers of the
3 resonance islands.

For $\gamma$ small enough, the resonances are well separated and the
motion 
is quasi-integrable. Chaos will develop when the resonances start
 to overlap. This is the celebrated Chirikov's overlap 
criterion~\cite{Chirikov79a}
 and its evaluation gives 
$\gamma\,\simeq \,0.1$ in our case. 
Thus chaos develops in phase-space regions where the kinetic energy term 
and the perturbation are of the same order of magnitude. Taking into 
account higher perturbation orders in~$\gamma$ 
will shift the position in phase-space
of the previous resonances
 as well as the frequency around their stable points. For instance, 
it can be seen in figure~\ref{fig:3yeux}-e
 that the stable island~$\mathcal{I}_+$ is centered on a point having a 
momentum slightly larger than~$+1$.
Perturbation terms of higher order will also introduce
 other resonances of smaller size. It is precisely the 
overlap of the infinite cascade of such resonances which gives rise to 
the chaotic layers. Nevertheless, Chirikov's criterion already gives 
a good order of magnitude for the onset of chaos.
For higher $\gamma,$ the previous three resonant islands of stability
have shrunk inside a large chaotic sea and eventually disappear
completely
(\textit{cf.}~figure~\ref{fig:3yeux}-f). Nevertheless
 revival of some stable islands can still be observed for some narrow
windows
of high~$\gamma$'s.
In our situation, chaos cannot invade the whole phase space 
but is bounded by regular coasts. This is so because chaos develops 
where resonances overlap. Sufficiently far away from the resonances,
atoms move 
so fast that they experience an average time-independent potential. 
Then chaos is 
absent and one recovers (quasi)-free motion when $|p|\gg1$.

\subsection{Typical classical phase space portrait in 
the chaos assisted tunnelling regime}
\begin{figure}[!ht]
\center
\psfig{file=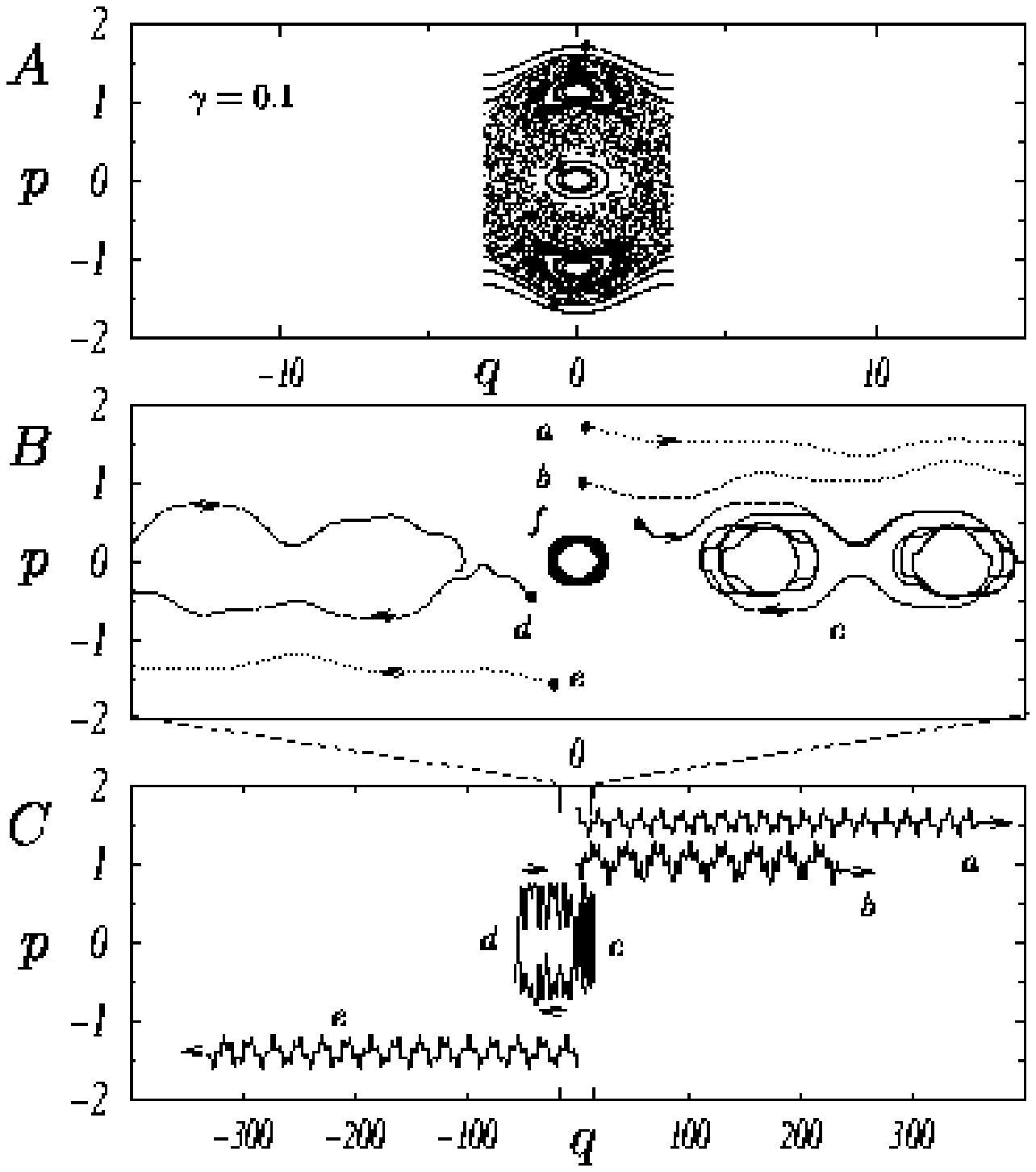}
\caption{\label{fig:3yeux_nonstrobo} Plots of some trajectories in phase
 space
for different initial conditions ($\gamma=0.1$). In $A,$
we display
the usual Poincar\'e surface of section like in figure~\ref{fig:3yeux},
that is 
stroboscopic plots which are fold according to the spatial 
periodicity. The small  black disks show some  initial conditions.
   In~$B$ 
and~$C,$ trajectories are plotted every $2\pi/100$ and,
unlike in~$A$, we let evolve $q(\tau)$ continously 
outside~$[-\pi,\pi[$.  
 Trajectories~$b$ and $f$  are trapped 
in the $\mathcal{I}_+$ and ${\mathcal{I}}_0$ resonance islands respectively.
Trajectories $a$ and~$e$ are  two examples of regular quasi-free 
motion.
$c$ and~$d$ correspond to chaotic motion, their initial conditions 
(at $\tau=0$) lie in the chaotic sea of figure~\ref{fig:3yeux}-d. 
        }
\end{figure}

The two resonant islands~$\mathcal{I}_{\pm}$,
 when they exist, are related by a
discrete symmetry: the time reversal invariance. 
As can be seen in figure~\ref{fig:3yeux_nonstrobo}, 
the atoms trapped 
in one island cannot classically 
escape from it: the boundaries of the islands play the role of a 
dynamical barrier which atoms cannot cross. 
Hence jumping from one island
to the other is a classically forbidden process
though it is expected to occur in quantum mechanics.
This is precisely the tunnelling situation we are interested in.
In fact,
 we will study the tunnelling
 between~$\mathcal{I}_+$ and $\mathcal{I}_-$ for~$\gamma$ varying from
0.1 to
 0.3 since, in that range, classical chaos may play a revealing role
though
the two stable islands still occupy a significant volume in phase space.

Note that, in the physical situation described by $H_{\mathrm{eff}}$, 
tunnelling occurs in
momentum coordinate instead of space coordinate as it is usually 
presented 
in standard textbooks. The denomination of 
``dynamical tunnelling''~\cite{Davis/Heller81a}, 
refers to this situation.
   The reason for investigating this situation is that manipulation 
of cold atoms allows for a better control (preparation and detection)
 of momentum rather than position.

\section{Quantum dynamics}\label{sec:quantumdynamics}

\subsection{Floquet-Bloch theory}
When an autonomous Hamiltonian is  spatially periodic, it is well 
known \cite{Ashcroft/Mermin76a} that its spectrum is organized in 
energy bands~$E_n(k)$. These bands are labelled by a set of integers, 
the band index~$n$, and depend continuously on a set of real numbers,
the Bloch numbers~$k$. 
As $E_{n}(k)$ and the associated eigenfunctions 
are periodic functions of the $k$'s, all the physical information
is contained in the first Brillouin zone. 
For 1D-systems, it is simply the interval
~$[-\frac{\pi}{Q},\frac{\pi}{Q}[$, where $Q$ is the spatial period 
of the Hamiltonian. 

When the Hamiltonian is time-periodic, with period $T$, 
the analogous of Bloch theory is Floquet 
theory~\cite{Floquet1883a,Cherry27a,Shirley65a}. The eigenvalues of the
 evolution operator~$U(\tau+T,\tau)$
 over one period take the form ${\mathrm{e}}^{-\imat\epsilon T/\hbare}$. 
The $\epsilon$'s are $\tau$-independent real quantities which are 
called the quasi-energies of the system. 
Due to the time-periodicity, the quasi-energy spectrum
as well
as the associated eigenfunctions
are now invariant under~$\epsilon\mapsto\epsilon+2\pi\hbare/T$.
 
For $H_{\mathrm{eff}}$, the application of Bloch and Floquet theorems
with $Q=T=2\pi$ yields a spectrum made of quasi-energy bands
$\epsilon_n(k)$ 
where $n$ goes over the whole 
set of integers (for a detailed derivation see
appendix~\ref{app:FBformalism}). 
For brievety, we will 
define~$|{n,k,\tau}\rangle$ the ket at time~$\tau$
 with Bloch angle~$k$, with 
quasienergy~$\epsilon_n(k)$ and
which is a solution of the Schrodinger's equation 
(following the
notations of 
appendix~\ref{app:FBformalism}, we 
have
set~$|{n,k,\tau}\rangle\DEF|{\psi_{\epsilon_n(k),k}(\tau)}\rangle$).
We will also define~$|{n,k}\rangle\DEF|{n,k,\tau=0}\rangle$. 

\begin{figure}[!ht]
\center
\psfig{file=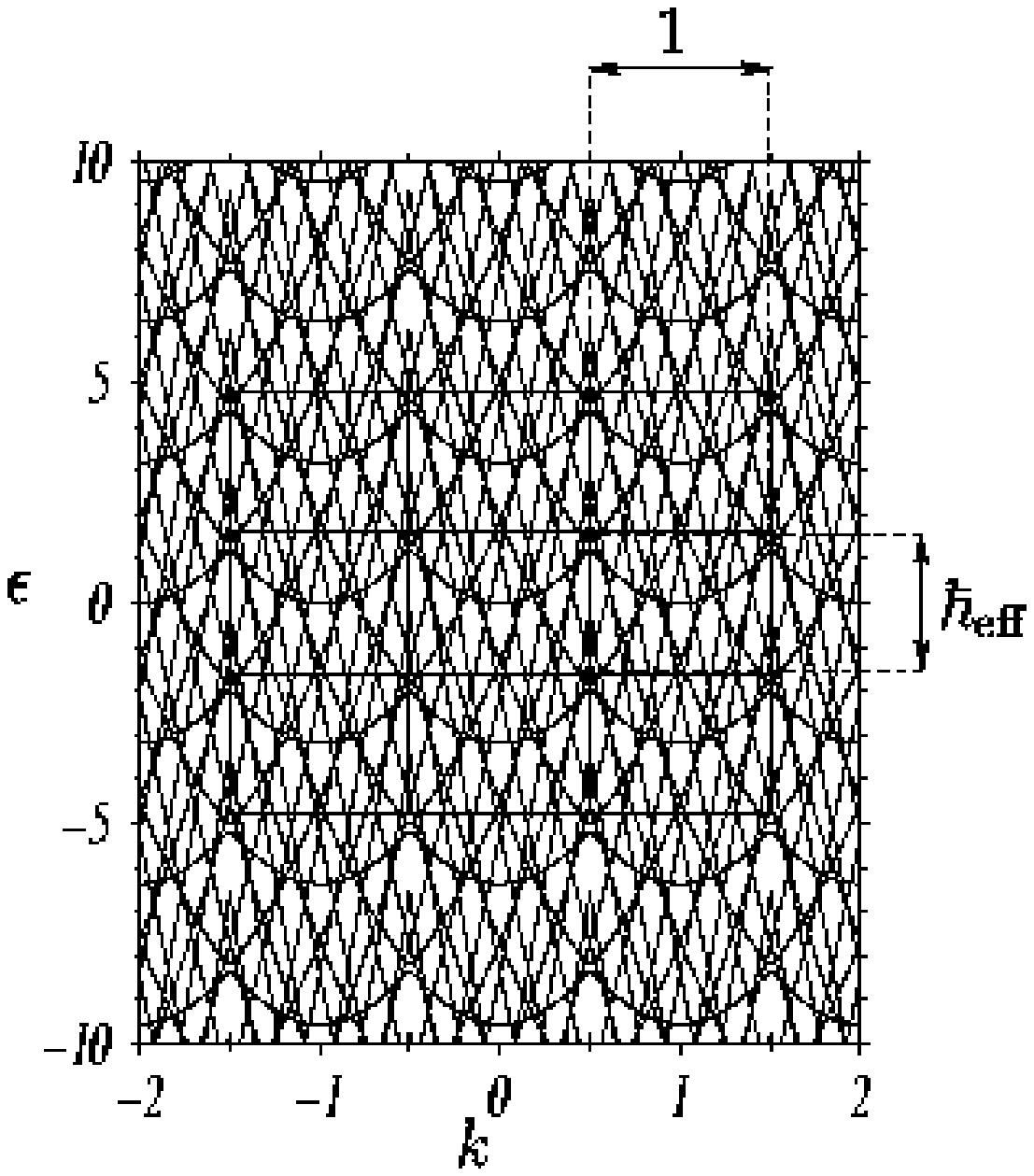}
\caption{\label{fig:biperiodicite} For a Hamiltonian which is 
 $Q$-periodic in space
 and $T$-periodic in time, the quasi-energy spectrum is made of bands
which are $2\pi/Q$-periodic in the Bloch numbers
 and $2\pi\hbare/T$-periodic
in the quasi-energies. Such a spectrum is shown here for
 Hamiltonian~\eqref{eq:ham_red} ($Q=2\pi$, $T=2\pi$) with $\gamma=0.18$ 
and $\hbare=3.1787$.}
\end{figure}

As it can be seen in figure~\ref{fig:biperiodicite}, the band spectrum
has
the topology of a torus since it is 
both periodic in quasi-energy (with period $\hbare$) and 
in Bloch number (with period $1$).

\subsection{Numerical calculation of the Floquet eigenstates}

As derived in appendix \ref{app:FBformalism}, the Floquet
eigenstates can be obtained by diagonalization of the
Floquet-Bloch operator, $\tilde{K}$ 
\begin{equation}\label{eq:Ktilde2}
\tilde{K}(\opp,\opq,\tau, k) = \frac{(\opp+\hbare k)^2}{2} -\gamma\cos
\opq\ (1 + \cos\tau )  
-\imat\hbare\frac{d}{d\tau}\;,
\end{equation}
with periodic boundary conditions both in time and space.
The eigenvalues, which depend on the Bloch vector $k$, 
are the quasi-energies of the system. The band spectrum is
symmetric with respect to the axis~$k=0$ since
the operator~\eqref{eq:Ktilde2}
 is invariant
under the transformation~$k\mapsto-k$ and~$p\mapsto-p$. 
From the expression of the Floquet-Bloch operator 
and the boundary conditions, it is very natural
to expand the eigenstates on a basis set composed of products
of the type $\phi_{lm}(\tau,q)=\exp(\imat n\tau)\exp(\imat m q)$ 
which automatically obey 
the periodic boundary conditions.
In such a basis, the operator  $\tilde{K}$ has very strong selection
rules,
namely:
\begin{equation}
|\Delta n| \leq 1\ \ \ {\mathrm and}\ \ \ |\Delta m| \leq 1
\end{equation}
All matrix elements violating one of these selection rules is
zero. Hence, the matrix representing the operator $\tilde{K}$  in this
basis is sparse and banded, and all matrix elements have simple
analytical expressions. This is well suited for numerical
diagonalization (powerful algorithms exist, for example the
Lanczos algorithm). All the numerical results presented
here use this method.  We checked
that the effect of the truncation of the basis  is negligible: 
the size of the
 Floquet matrix is considered as sufficiently large
 when increasing it modifies the value of 
the quasi-energy on the scale of the  numerical noise only, say
$10^{-15}$
 in double precision. 
Not only this criterion is a proof of algorithmical
 convergence but also
it is a safeguard against numerical  discrepancies  since we are
 looking for exponential small quantities. 

\subsection{Husimi representation and classification of the quantum
states}
Classical dynamics is very illuminating when describing 
 the states~$|{n,k,\tau}\rangle$. In order to strengthen 
the correspondence between classical phase-space structures
 and quantum states, it is convenient to work with the Husimi 
representation of quantum states~\cite{Louisell73a}.

Such a representation associates to each quantum state 
$|{\psi}\rangle$ a phase space 
function~$\psi^{\mathrm{\scriptscriptstyle H}}(p,q)$ (where $p$ and 
$q$ are real numbers) defined by
\begin{equation}
\psi^{\mathrm{\scriptscriptstyle H}}(p,q) \DEF N_\psi\;|\langle{z}|{\psi}\rangle|^2        
\end{equation}
where~$|{z}\rangle$ is the normalized coherent state corresponding to
the 
complex number~$z=(q+\imat p)/\sqrt{2\hbare}$. $N_\psi$ is a 
$(p,q)$-independent normalization factor. Because~$|{z}\rangle$
is a minimal gaussian wave packet with average momentum $p$ and 
average position $q$, the Husimi function 
$\psi^{\mathrm{\scriptscriptstyle H}}(p,q)$
contains some information about the degree of localization 
of~$|{\psi}\rangle$ in phase space.

\begin{figure}[!ht]
\center
\psfig{file=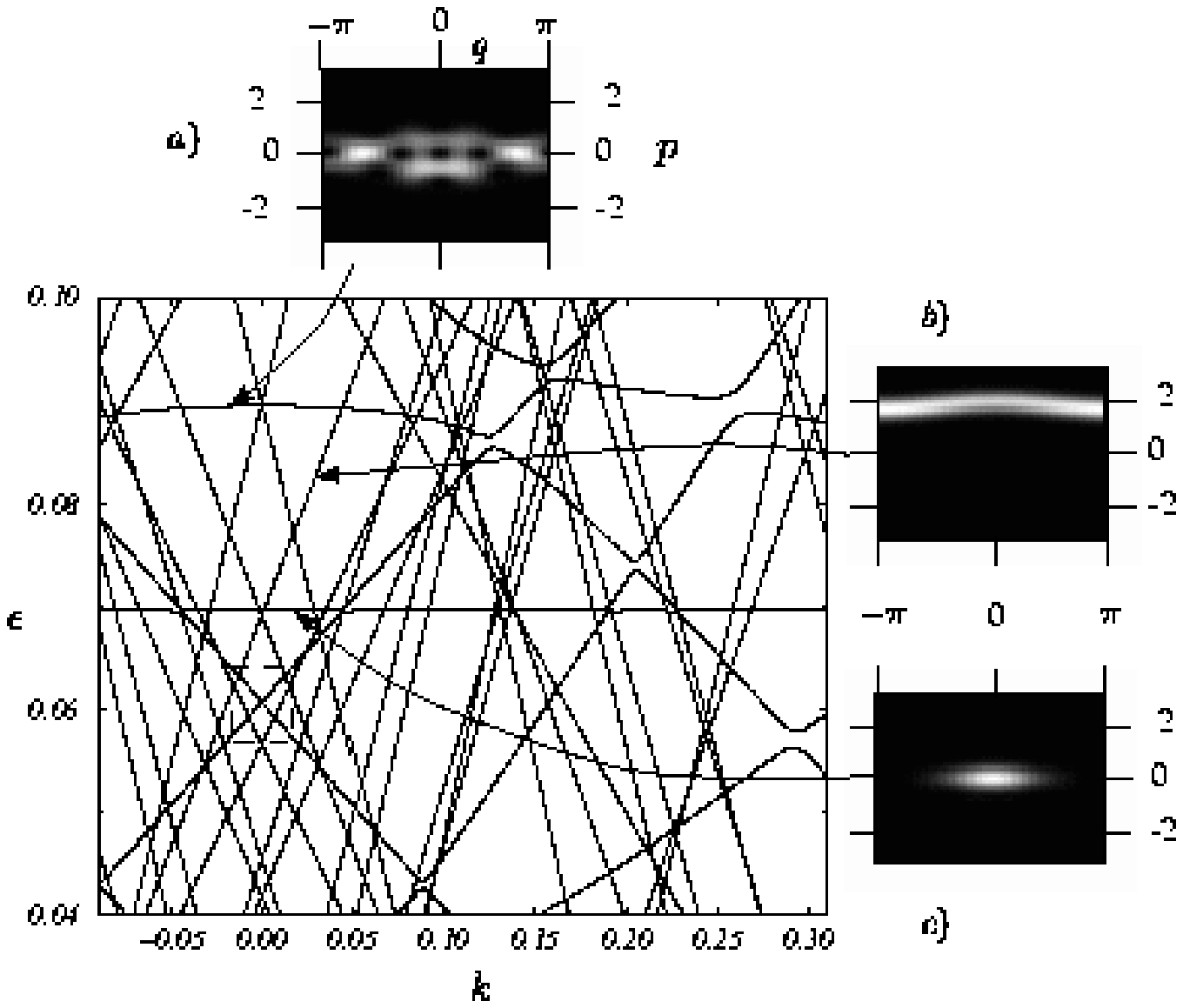}
\caption{\label{fig:bandes_un} Quasi-energy bands for~$\gamma=0.18$ 
and~$\hbare=0.2037$ and some Husimi representations of  typical states. 
In~$b,$ we
show a quasi-free state with a well defined average velocity 
(the derivative of the energy level with respect to $k$)
localised in phase space on regular trajectories 
(compare with figure~\ref{fig:3yeux}-e). On this
scale, 
the avoided crossings
with other bands cannot be resolved.
In~$c,$ we show a state 
localised in the central stable island~${\mathcal{I}}_0$ 
(actually the ``ground state'').
 Far from quasi-degeneracies
the average velocity of this state is zero.
In~$a,$  we give an example of a chaotic state, 
 whose Husimi function is
localised in the chaotic sea (compare with figure~\ref{fig:3yeux}-e). 
Unlike the former states, 
the average velocity 
is fluctuating when varying the Bloch angle~$k$. 
The band spectrum is
symmetric with respect to the axis~$k=0$ since
the operator~\eqref{eq:Ktilde2}
 is invariant
under the transformation~$k\mapsto-k$ and~$\opp\mapsto-\opp$. 
The tunnelling 
situation
due to the time reversal symmetry corresponds to the dashed squared zone 
(around~$k=0$) which is enlarged
in figure~\ref{fig:bandes_deux}.
        }
\end{figure}

\begin{figure}[!ht]
\center
\psfig{file=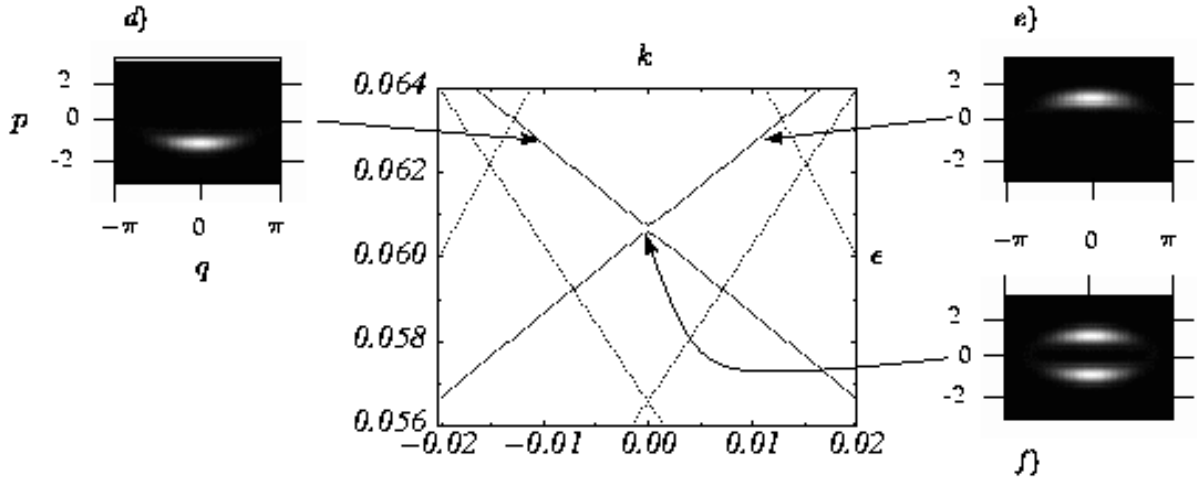,width=18cm}
\caption{\label{fig:bandes_deux} Quasi-energy bands for~$\gamma=0.18$ 
and~$\hbare=0.2037$ and some Husimi representations of  typical states.
It is a zoom of the dashed squared zone in figure~\ref{fig:bandes_un}.
For~$k\neq0$, one can find states, like in~$d$ or~$e,$ 
whose Husimi function is
 localised  in one stable island~${\mathcal{I}}_-$ 
or~${\mathcal{I}}_+$ (compare with figure~\ref{fig:3yeux}-e). 
In the frame where the center of the 
island is fixed, these states correspond to the ``ground'' states 
(some  excited states may of course exist if~$\hbare$ is small 
enough as can be seen in figures 
\ref{fig:suiviadiabatique} 
and~\ref{fig:suiviadiabatique2}).
 For~$k=0$, we recover time reversal 
symmetry through the existence of quasidegenerate doublets of symmetric 
or antisymmetric
combinations. This is the typical tunnelling situation: following 
adiabatically with~$k$
the state in~$e$ (which has an average velocity
about~$+1$), by decreasing~$k$
we get the state in~$d$ which has a reversed velocity. 
This reversal of the velocity
is a classically forbidden process (compare with orbit~$b$ in 
figure~\ref{fig:3yeux_nonstrobo}). 
        }

\end{figure}

The minimal cell size in phase-space allowed by the Heisenberg
inequalities 
is $\hbare$. Let us see how classical phase space structures of typical
size
larger than~$\hbare$ are mirrored
at the quantum level.
In figure~\ref{fig:bandes_un} and~\ref{fig:bandes_deux} we have plotted 
some of the~$\epsilon_n(k)$ 
corresponding to the Hamiltonian~\eqref{eq:ham_red} for specified 
fixed values of~$\gamma$ and~$\hbare$. Von Neumann and Wigner 
arguments~\cite{vonNeumann/Wigner29a} claim that, 
generically, no exact degeneracy can occur: one rather gets 
avoided crossings.
Of course, this is relevant provided the minimal energy splitting
is greater than the  resolution in energy. 
 Some 
Husimi
functions 
are plotted in
 figure~\ref{fig:bandes_un} and~\ref{fig:bandes_deux} ($a$ to~$f$)
\footnote{A technicality should be mentioned here:
$|{u_{\epsilon,k}}\rangle$
        and~$|{\psi_{\epsilon,k}}\rangle$ obey some spatial boundary 
        conditions
        which are lost when working with their
         Husimi representations, essentially
         because the coherent
         states do not fulfill themselves these properties. 
        To deal with spatially (quasi-)periodic phase-space functions
        it is necessary to unfold the coherent
states\cite{Leboeuf/Voros92a} into
        \begin{equation}
                |{z,k}\rangle\DEF\sum_{m\in{\mathbf{Z}}}
                        {\mathrm{e}}^{\imat mkQ}|{z+mQ}\rangle
        \end{equation}
        and define for instance
        \begin{equation}
                \psi^{\mathrm{\scriptscriptstyle H}}_{n,k}(p,q,\tau)
                \DEF N\;
                |\langle{z,k}|{\psi_{n,k}(\tau)}\rangle|^2\;
        \end{equation}
         }.

In appendix~\ref{app:FBformalism}, it is
shown that we have 
\begin{equation}
\label{eq:averagev}
v_{n,k}\DEF\frac{1}{T}\int_0^T\langle{n,k,\tau}|\,\opp\,|{n,k,\tau}\rangle\,d\tau
=\frac{1}{\hbare} \frac{\partial \epsilon_n}{\partial k}\;, 
\end{equation}
 which
generalizes the velocity theorem~\cite[Appendix E]{Ashcroft/Mermin76a}
to
periodic time-dependent hamiltonians.

Figure~\ref{fig:bandes_un}-b shows an example of 
the Husimi representation of
a state with sufficiently high average velocity
to be alike a free eigenstate of~$H_0$. Far from quasi-degeneracies 
it is localized in a narrow 
strip of width~$\Delta p\sim2\pi\hbare/\Delta q\sim\hbare$ (since~$\Delta q$
covers~$2\pi$) and which is
centered on one of the two classical phase space trajectories of 
energy about~$v_{n,k}^2/2$ (compare with figure~\ref{fig:3yeux}-e).
Its quasi-energy band 
(figure~\ref{fig:bandes_un}) is an arc of the parabola of the free
motion but
can hardly be distinguished from a straight line of slope~$v_{n,k}$
if $k$ is restricted to one Brillouin zone. 
We will naturally call these states quasi-free states.

Some states have their Husimi functions localized in the resonant stable
islands (in ${\mathcal{I}}_0$ but also in~${\mathcal{I}}_\pm$).
 The number of these
states is semiclassically given by the volume of these islands divided 
by~$2\pi\hbare$. 
Far away from quasi-degeneracies,
 these states are at any time centered 
 on the 
stable  periodic orbit: it can be explained within a semiclassical
approach 
and can
be observed in figure~\ref{fig:bandes_un}-c for a state localized 
in~${\mathcal{I}}_0$ and 
in figures~\ref{fig:bandes_deux}-d,e for states in~${\mathcal{I}}_-$, 
${\mathcal{I}}_+$ respectively.
 Their average velocity as well as their 
Husimi functions depend exponentially weakly on the Bloch parameter~$k$.

The last class of 
states which can be encountered corresponds to chaotic ones, that is
to say states whose Husimi functions are negligible on a typical
distance 
of~$\sqrt{\hbare}$ out of the chaotic
seas (\textit{cf.}~figure~\ref{fig:bandes_un}-a).
 Unlike the previous ones, their Husimi functions
 are very sensitive to any variation 
of~$k$ since they are delocalised in the whole chaotic sea which
spreads over all elementary  cells. The large classical distribution
of possible velocities is to be linked to the very fluctuating slopes of
the
quasi-energies as~$k$ is varied.

\subsection{Tunnelling states}\label{subsec:tunnellingstates}

Although the system as a whole is of course
time-reversal invariant, this is no longer true for its restriction
at a fixed value $k$ of the Bloch vector. Indeed, the operator
$\tilde{K}$ is not time-reversal invariant, because
of the crossed term $k\opp .$ In other words, the time-reversed partner
of a state with Bloch vector $k$ is a state with Bloch vector $-k.$
It is only at the special value $k=0$ (and also $k=1/2$ since $k$
is defined modulo 1)
that  $\tilde{K}$ is invariant under time reversal symmetry.
Therefore it corresponds to the typical situation of tunnelling 
between~$\mathcal{I}_{+}$ and~$\mathcal{I}_{-}$. Every island state
localised about~$|p|=1$ is quasi-degenerate with another one. These
doublets
represent 
a symmetric and an antisymmetric combination 
of states  localised in one island only 
(\textit{cf.}~figure~\ref{fig:bandes_deux}-f).
 The 
energy splitting $\Delta\epsilon_n$
 of these states for~$k=0$ is precisely the signature of tunnelling: 
it is $\pi\hbare$  
divided by the typical time an atom takes 
to jump from one island to its time-reversed image, \textit{i.e.}~to
reverse 
the sign of its velocity.

\section{Chaos assisted tunnelling}\label{sec:cat}

\subsection{Large fluctuations}

\begin{figure}[!ht]
\center
\psfig{file=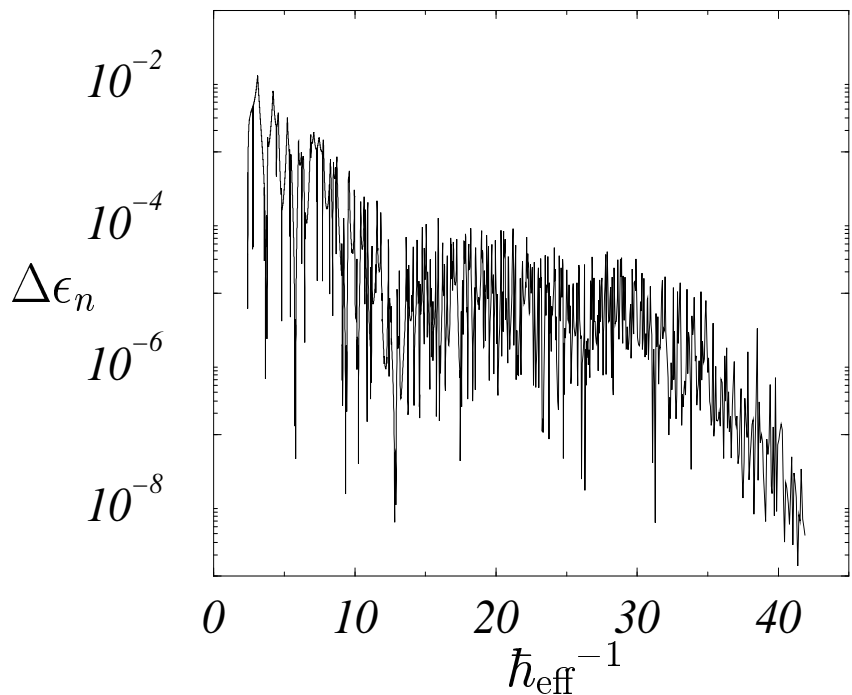}
\caption{\label{fig:fluctuations_hbar} Fluctuations of the 
energy splittings~$\Delta\epsilon_n$ between pairs of
symmetric/antisymmetric
states localized in the~$\mathcal{I}_{\pm}$ resonance islands (here
shown
for the ``ground state" inside the island).
The classical dynamics is fixed at~$\gamma=0.18$ 
(compare with figure~\ref{fig:3yeux}-e). The existence 
of large fluctuations
 over several orders of magnitude is a signature of chaos assisted
tunnelling.
On the average, $\ln|\Delta_{\epsilon_n}|$
appears to decrease more or less linearily with~$\hbare^{-1}$
 except for
 the plateau at $15\le\hbare^{-1}\protect\le30$,  
        }
\end{figure}

\begin{figure}[!ht]
\center
\psfig{file=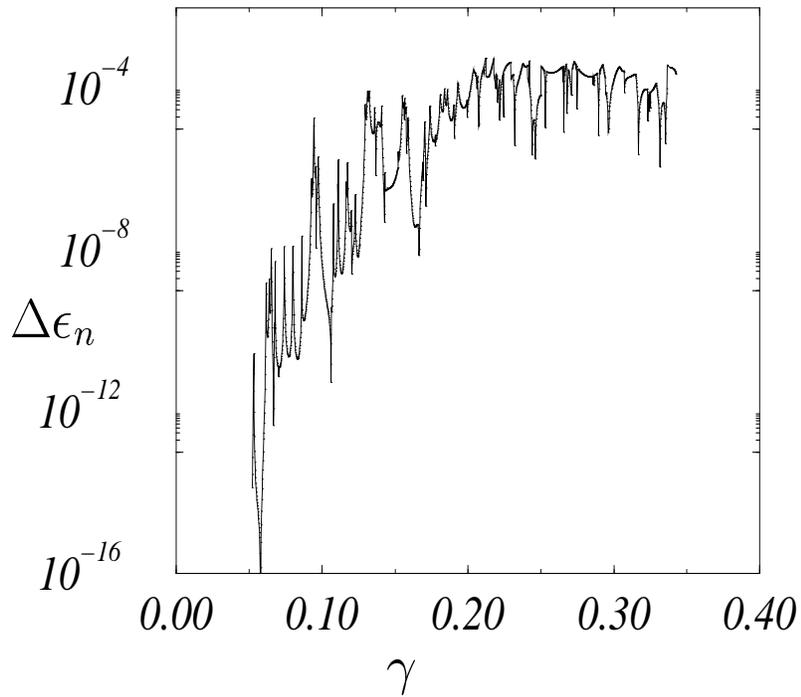}
\caption{\label{fig:fluctuations_gamma} Fluctuations of the 
energy splittings~$\Delta\epsilon_n$ between the pair
of symmetric/antisymmetric states  
($\hbare^{-1}$ is fixed at~$19.309$) as a function of
$\gamma.$  Again,
the large fluctuations
 over several orders of magnitude are a signature of chaos assisted
tunnelling. The global increase with $\gamma$ is due to the growth
of the chaotic sea as $\gamma$ increases, see figure~\ref{fig:3yeux}.
        }
\end{figure}

After having selected the two quasi-energy bands corresponding to the 
two states which are
localized  the more deeply inside the islands~$\mathcal{I}_{\pm}$,
we are able to plot the splitting as a function of~$\hbare$.  
The great advantage of studying fluctuations when the 
effective Planck constant is varied is that it does not affect 
the classical dynamics. The behavior of the splitting is very
different  whether
chaos is present at the~$\hbare$ scale or not.
In the chaotic regime (\textit{cf.}~figure~\ref{fig:fluctuations_hbar} 
and~\ref{fig:fluctuations_gamma}), that is
when~$\hbare$ varies in a range where chaotic seas
 can be resolved, 
the splittings vary rapidly versus the change of any parameter, in our
case $\hbare.$ Moreover, the variations of the splittings, despite 
being perfectly deterministic, are apparently
erratic -- without any regular structure -- and cover
several orders of magnitude. They show that direct coupling to the chaotic 
sea is the key 
mechanism for their understanding and are a signature of chaos assisted 
tunnelling\cite{Zanardi+95a,Zakrzewski+98a}. In fact these huge fluctuations are 
reminiscent of the universal conductance fluctuations observed in 
mesoscopic systems \cite{Washburg/Webb86a,Feng/Lee91a}
 since 
tunnelling is nothing else than wave transport from one stability 
island to the other. 
In contrast,
in the regular regime where chaotic seas
are smaller than~$\hbare$, the splittings are expected 
to vary smoothly\cite{Wilkinson86b}.

In figure~\ref{fig:fluctuations_hbar}, we show the splittings of the
pair of
states
localized at the center of the resonances islands $\mathcal{I}_{\pm},$
as a 
function of $\hbare.$ They display huge fluctuations over about
4 orders of magnitude while the general trend is a decrease as $\hbare
\to 0.$
Similarly, when plotted as a function of $\gamma,$ 
see figure \ref{fig:fluctuations_gamma},
they also display large fluctuations. The general trend is here a fast
increase of the typical splitting with $\gamma ;$ this is associated with
the
shrinking of the regular island when $\gamma$ increases which results
in a increasingly large tunnelling probability
The overlaps of the  regular states (still supported by the islands)
with the chaotic states increases. Therefore the  coupling between
 the two components 
of the tunnelling doublets which involves the chaotic states increases
 as well.
 
In order to understand both the general trend and the origin of the
fluctuations, two points of view can be used: a quantum point of view 
and a semiclassical one.

\subsection{Semiclassical interpretation}
If the chaotic sea is large, it is rather intuitive that it can be
easier to 
first tunnel from the center of the regular island to the chaotic sea, 
then propagate
freely in the chaotic sea to the vicinity of the symmetric island and
finally tunnel to the center of the symmetric island than directly
tunnelling
between the two islands. The crucial point is that because the chaotic 
component
is explored rapidly and densely, it does not cost anything to cross the
chaotic sea. Tunnelling trajectories can be viewed as complex
trajectories 
(i.e.
with complex position and momentum) which are real at both the starting
(in the initial island) and ending (in the symmetric island) points.
The tunnelling amplitude associated with a single tunnelling orbit
is essentially $\exp (-\mathrm{Im}(S)/\hbare)$ 
where $S$ is the complex action of
the tunnelling orbit. In a usual 1-dimensional system (like a
double-well),
there is only one such trajectory at each energy and the tunnelling rate
thus displays the well-known exponential decrease. In a chaotic system,
it may happen that there is a whole set of tunnelling trajectories whose
imaginary parts of the action are essentially identical. In such
conditions, 
the
actual tunnelling amplitude is the sum of all individual amplitudes (each
taken with its proper phase) which results in a very complicated
quantity
which fluctuates when parameters are changed. In some sense, this is
analogous
to the speckle pattern obtained when plenty of optical rays with various
geometries are randomly interfering. This is the very origin of the
(deterministic) fluctuations of the tunnelling rates and
consequently of the energy splittings. The general trend (exponential
decrease) is related to the typical imaginary part of the action of
the tunnelling trajectories.

\subsection{Quantum point of view}
A complementary quantum point of view is possible. One can divide the
eigenstates of the system in two subsets: the ``regular" states
localized
in the resonance islands and the ``chaotic" states localized in the
chaotic sea.
The two sets are only weakly coupled by tunnelling. Because there are 
two symmetric 
islands, the regular states are essentially doubly degenerate
(neglecting 
{\em direct}
tunnelling). The chaotic sea also has the two-fold symmetry and states
can be
classified as even or odd. The two series of odd and even states are 
ignoring each other.
Hence, when by accident, a even chaotic state is almost  degenerate with
a even regular state, they repell each other: at the same time, there 
is usually
no odd chaotic state with the same energy. Thus, the odd regular state
is
not significantly repelled. Hence, the splitting appears because
of different shifts of the even and odd regular states. Close to any
avoided crossing between either the odd or the even regular state and
a corresponding chaotic state, a large splitting is obtained.
On the contrary, far from any avoided crossing, the splitting is 
small. Hence, the fluctuations are just associated with the
existence of a large number of successive avoided crossings. The typical
size of these fluctuations is related to the typical size
of the avoided crossings while the typical parameter range of these
fluctuations is the distance (in parameter space) between two
consecutive 
avoided crossings. A model implementing this idea (each regular state 
is independently and randomly 
coupled to the chaotic states of the same symmetry) has been
proposed in~\cite{Leyvraz/Ullmo96a} and further used in \cite{Zakrzewski/Delande93a}.
In this model, the chaotic states are modeled by a Hamiltonian
belonging to the Gaussian Orthogonal Ensemble of random matrices
while the coupling between the regular state and the chaotic
state is also taken as a random Gaussian variable.
With these assumptions, the splitting distribution can be
calculated. Let  us denote by $\Delta$  the mean level spacing between
chaotic states and by $\sigma$ the typical strength of the coupling between
the regular states and the chaotic sea. Only if $\sigma \ll \Delta$
is the regular state weakly coupled to the continuum (if this inequality
is
violated, the regular state is completely diluted in the chaotic sea
by the strength of the coupling). We thus assume the inequality
to be valid. Then, the distribution of splittings 
$\Delta\epsilon$ is
given by:
\begin{equation}
\left\{
\begin{array}{l}
\displaystyle P(\Delta \epsilon) = \frac{1}{\pi} \frac{s}{s^2+\Delta\epsilon^2}
\ \ \ {\mathrm for}\ \ |\Delta\epsilon|<\sigma \\
\displaystyle P(\Delta \epsilon) \simeq 0 \ \ \ {\mathrm for}\ \ |\Delta \epsilon| >
\sigma
\end{array}
\right.
\end{equation}
where
\begin{equation}
s = \frac{\sqrt{2} \pi \sigma^2}{\Delta}
\end{equation}
The interpretation is rather simple. The maximum splitting is observed
exactly
at the avoided crossing where the levels are shifted by $\pm\sigma/2$ on
both sides
of their unpertubed positions. Hence, the splitting cannot be larger
than about $\sigma.$
In fact, there is an exponentially decreasing tail in the distribution
$P(\Delta\epsilon)$
(associated with the Gaussian fluctuations of $\sigma$)
which we do not detail here because it is not relevant in our present
case. 
$s$ is the typical splitting one expects to observe: it corresponds to
the
shift typically due to the closest chaotic state. The full distribution
is
a (truncated) Cauchy distribution: it is obtained as the overall effect
of all chaotic states lying above or below in energy.
Note that, in the absence of the truncation of the Cauchy distribution,
the average splitting is not defined because the corresponding integral
diverges. Hence, it is better to discuss the typical splitting $s$
rather
than the average splitting. It is the slow decrease of the Cauchy
distribution for large splitting which is responsible for the huge
fluctuations
in the splittings, which can be as large as $\sigma \gg s.$ This is
reminiscent of random processes such as L\'evy flights where rare events 
are dominant\cite{Bouchaud/Georges90a}.

\begin{figure}[!ht]
\center
\psfig{file=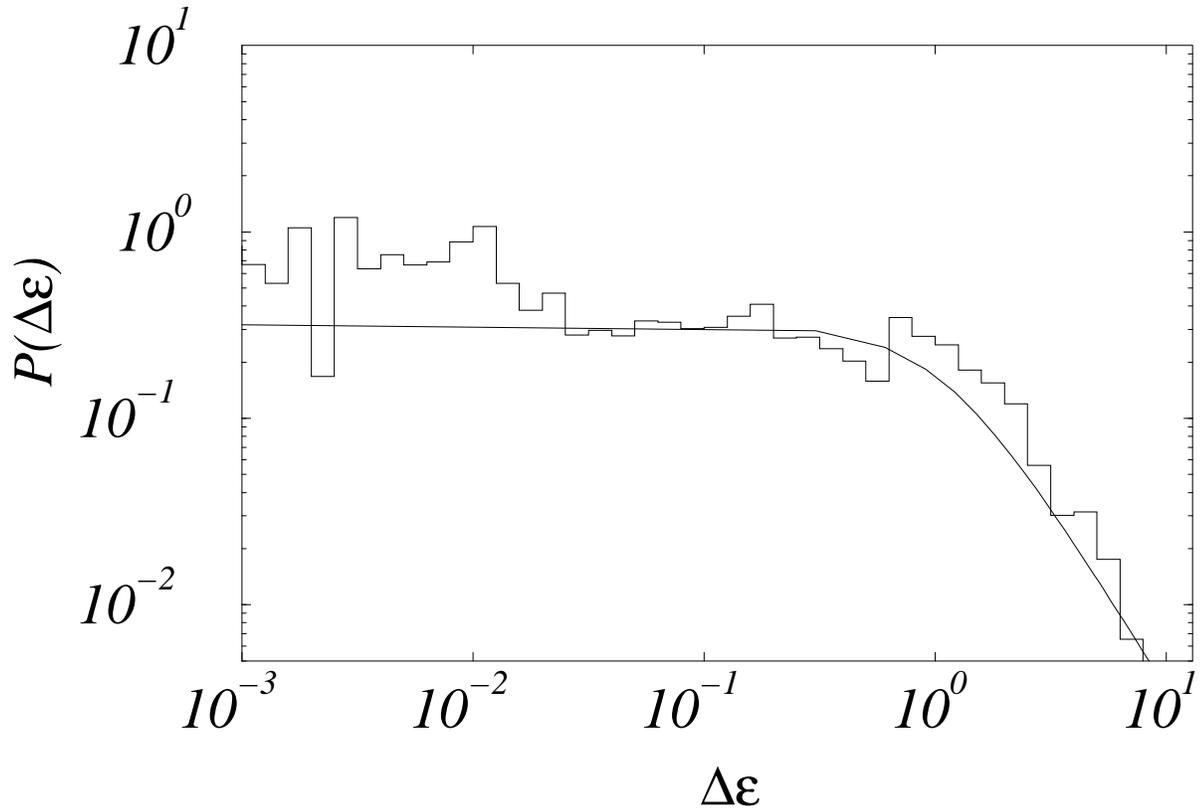,width=16cm,angle=-90}
\caption{\label{fig:LUlaw} Statistical distribution of the
energy splittings $\Delta\epsilon$ 
(normalized to the typical splitting) between pairs of
symmetric/antisymmetric states localized inside the~$\mathcal{I}_{\pm}$
resonance
islands ($\gamma=0.18$, $k=0),$ represented on a double logarithmic
scale.
On can clearly distinguish two regimes: constant at small  
$\Delta\epsilon$
followed by a $1/\Delta\epsilon^2$ decrease and finally a rapid cut-off
(not shown in the figure). The solid line is the Cauchy distribution
predicted
by Random Matrix Theory.} 
\end{figure}

In fig.~\ref{fig:LUlaw}, we show the statistical distribution of
splittings
that we obtain numerically in the chaotic regime (normalized to the typical splitting in order
to work with~$s=1$). The distribution is
shown
on a double logarithmic scale and compared with the Cauchy distribution. 
On can clearly see two regimes: for small $|\Delta\epsilon|,$
$P(|\Delta\epsilon|)$
is almost constant and decreases with a slope -2 for large
$|\Delta\epsilon|.$
The agreement with the Cauchy distribution is very good, which 
proves that the model catches the essential part of the physics in this
system.

The typical splitting $s$ is proportional to the square of the tunnelling
matrix element from the initial state to the chaotic sea. Hence, it is
expected
to decrease roughly like $\exp(-2\mathrm{Im}(S)/\hbare)$ where $S$ is the complex
action of tunnelling orbits (the mean level spacing scales
as a power of $\hbare$ and is thus a correction to the main exponential
decrease). This is roughly what is observed in
fig.~\ref{fig:fluctuations_hbar}.
Note however that there is a plateau in the range 
$15\le\hbare^{-1}\le30.$
A similar observation has been done in~\cite[fig. 3]{Roncaglia+94a}.
Although this is not a crucial problem (the statistics of the 
splitting distribution is not affected), a detailed explanation of this
behaviour is still lacking, see however~\cite{Roncaglia+94a}.
Finally, it should be interesting to calculate explicitely some
of the complex tunnelling orbits in our specific system in order
to compare the imaginary part of their actions with the slope
in fig.~\ref{fig:fluctuations_hbar}. This work is currently under
progress.

\section{Experimental signal}\label{sec:experimentalsignal}

As it can be seen in figure~\ref{fig:bandes_un}, the splittings we want 
to measure
correspond to very tiny scales among the other structures in the band
spectrum. 
For say~$\hbare\simeq0.1$, the tunnelling times are about 
$\hbare/\Delta\epsilon\sim10^3$ 
times the typical period (T$\sim \mathrm{few} \mu\mathrm{s}$). The observation of
atoms
having reversed their velocity  is therefore still possible since 
during $1\ \mathrm{ms}$
spontaneous emission has not begun to spoil our dynamical model.
Nevertheless, 
measuring tiny splittings which are hidden so deeply in the spectrum 
is far from 
being straightforward. 
Several steps are needed: preparation of the initial state, the
experiment
itself (where chaos assisted tunnelling takes place) and the analysis of
the
final state. During the first step, one must prepare a state
localized inside
one resonance island, i.e. localized both in position and momentum
spaces.  The idea is to use an adiabatic transfer from an initial state
extended in position space by slowly branching the effective potential.
Although the second step looks trivial (one just has to wait),
the dispersion in the vector angle $k$ makes things much more difficult
as only the $k=0$ states are related by time-reversal symmetry (see
section~\ref{subsec:tunnellingstates}) and thus tunnel relatively fast. It is however possible
to overcome this difficulty as explained below.
Finally, the detection should be rather easy, using velocity dependent
Raman transitions. We now explain in detail how the various steps
can be worked out.

\subsection{Adiabatic preparation of the atoms in one
 lateral stable island}

The first step consists in preparing an initial cloud of cold Rubidium
atoms
in order to have it 
located in one of the stable island, say~$\mathcal{I}_{+}$, only.
Using a standard magneto-optical trap, one can obtain a more or less
thermal distribution of atoms with velocity of the
order of few times the recoil velocity 
$v_{\mathrm rec}=\hbar k_L/M=6~\mathrm{mm/s}$. 
However -- as shown below --
this is probably too much for a good measurement of the
tunnelling splitting. Additional techniques (side-band
 cooling\cite{Morinaga+99a}
, Raman cooling\cite{Kasevich/Chu92a,Reichel+95a}) make it possible to obtain
a sub-recoil velocity distribution, i.e. atoms 
with an average momentum 
$p_0$
and a thermal dispersion~$\Delta p_x=M\Delta v=M \alpha
v_{\mathrm{rec}}$
with $\alpha$ significantly smaller than 1.
We chose the initial momentum to be $M\delta\omega/2k_L$ so that, on
average, the atoms exactly follow one of the sliding standing
wave created by a pair $\omega_{\pm}$ of laser beams.

The next step is to slowly (i.e. adiabatically) switch on
the standing waves. During this phase, the spatial periodicity
is preserved, and the Bloch vector $k$ is thus a conserved quantity. 
Initially, the momentum $p_x$ is nothing but
the Bloch vector (modulo a integer multiple of the
recoil momentum). Thus, by preparing a sub-recoil initial state,
one populates only a small range of $k$ values and, for each 
$k$ value populated, a single state (momentum eigenstate). 
Otherwise stated, the initial momentum distribution has become a 
statistical mixture of Bloch states with $\Delta k=\alpha/2$ 
in a single energy band. 
More generally, if the initial momentum distribution is not
a sub-recoil one, but has a width equal to $\alpha$ recoil momenta
(with $\alpha > 1$),
about $\alpha$ bands will be populated.

\begin{figure}[!ht]
\center
\psfig{file=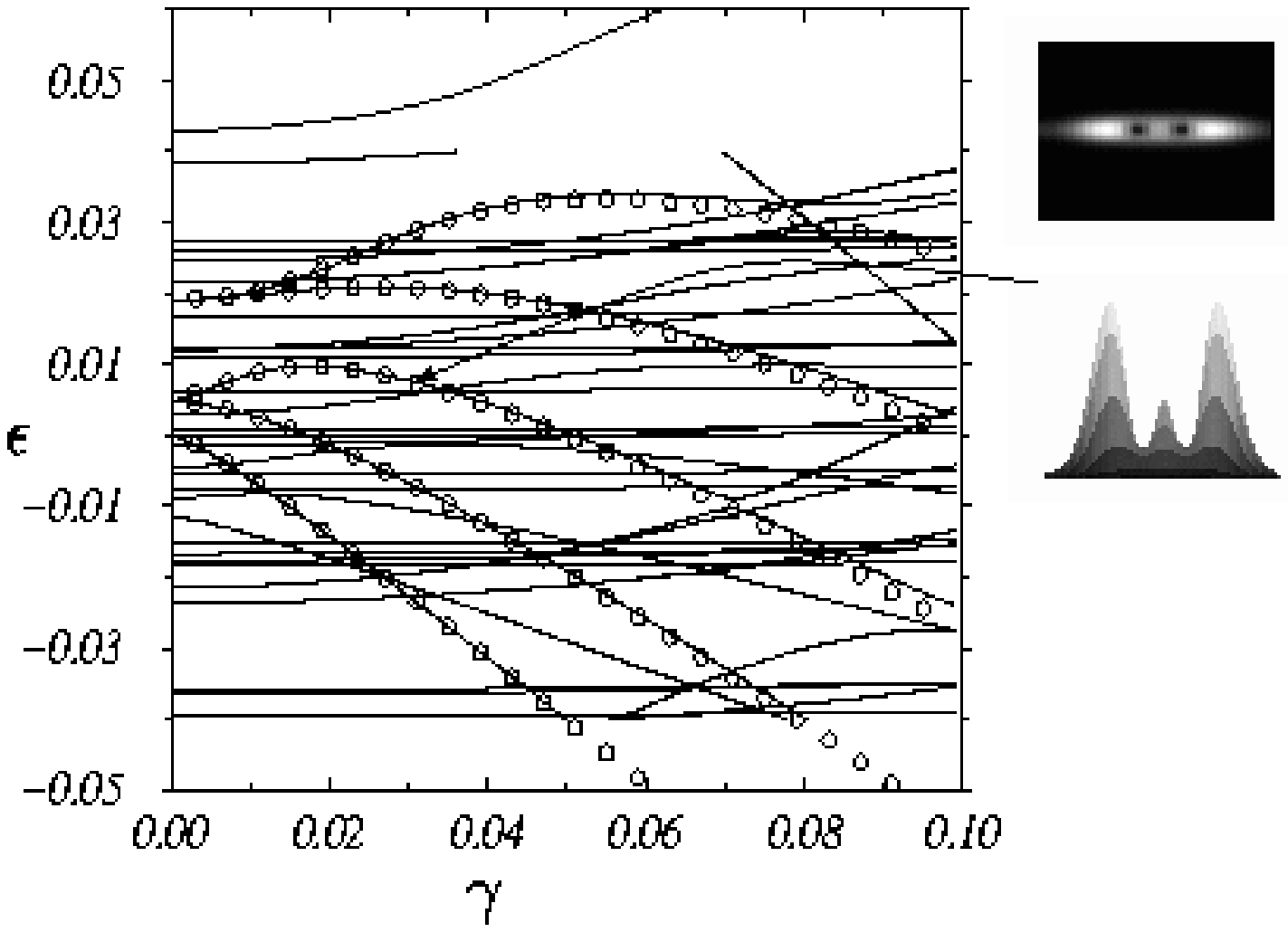}
\caption{\label{fig:approxpendule} Comparison between the quasi-energies
of Hamiltonian~\eqref{eq:ham_red} (lines) and the energies (circles) 
obtained from  the pendulum
approximation in the stable island~${\mathcal{I}}_0$ 
($\hbar=.0976,k=0$). The avoided crossings  which  appear
at~$\gamma\sim0.07$ illustrate the  influence of  classical narrow
chaotic seas
on the quantum properties. The Husimi distribution of the second
excited state localised in~${\mathcal{I}}_0$ for~$\gamma=0.03$ is
shown.
}
\end{figure}

\begin{figure}[!ht]
\center
\psfig{file=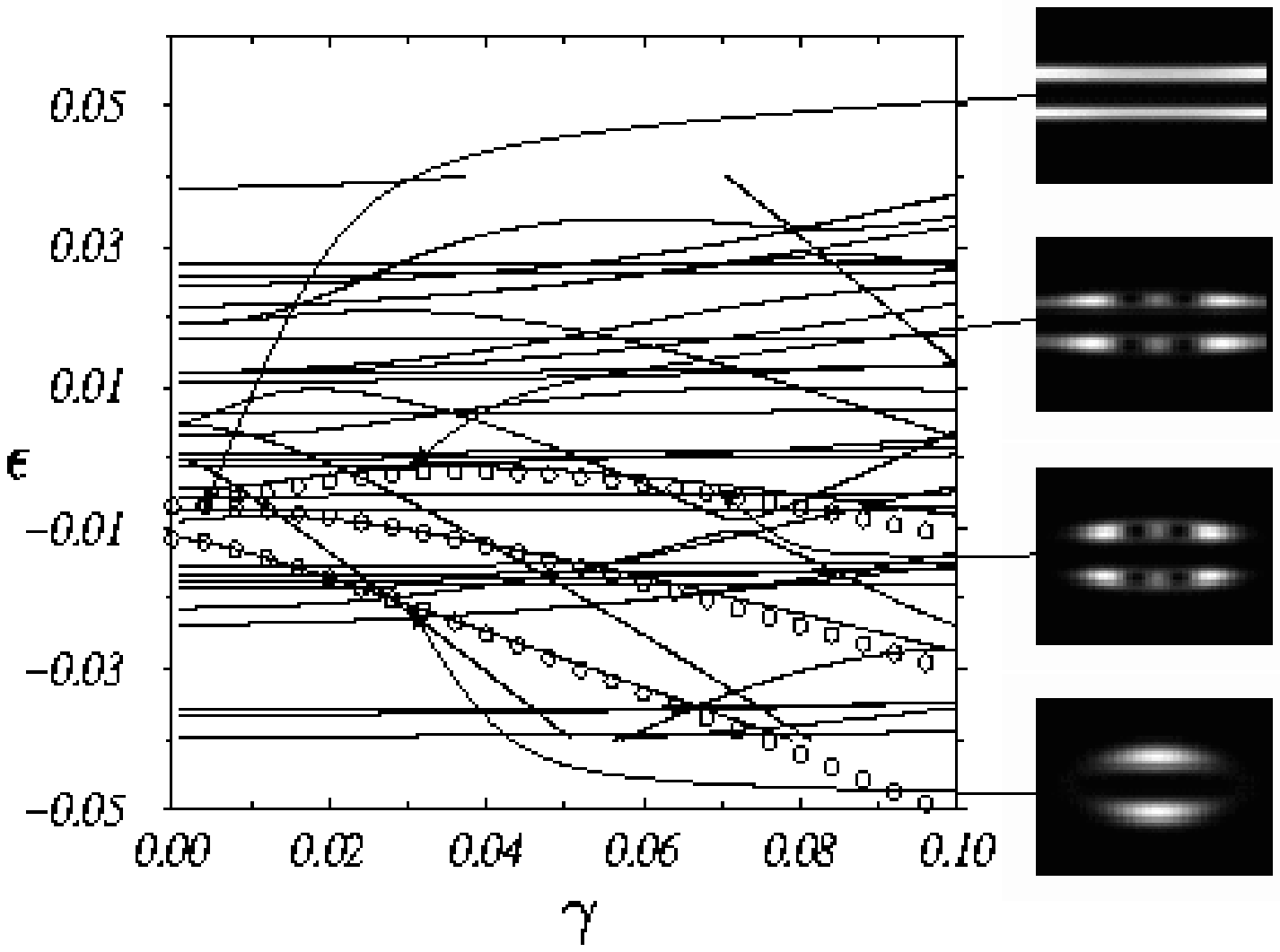}
\caption{\label{fig:approxpendulebis} Comparison between the
quasi-energies
of Hamiltonian~\eqref{eq:ham_red} (lines) and the energies (circles) 
obtained from  the pendulum
approximation in the stable islands~${\mathcal{I}}_\pm$  
($\hbar=.0976,k=0$). The three upper Husimi plots show how a 
quasi-free state doublet becomes progressively localised in the stable
island when
$\gamma$ increases.
The lower 
Husimi plot  corresponds to the ``ground'' state
 of the pendulum approximation
in the stable islands~${\mathcal{I}}_\pm$  for~$\gamma=0.03$.
}
\end{figure}

Switching on the standing waves increases the optical potential 
$V_0$ and therefore
$\gamma$ from zero and  enlarges the
resonance islands. In the sliding frame moving
 with velocity~$p_0/M=\delta\omega/2k_L$,
the atoms feel a pendulum like potential~$-\frac{1}{2}\gamma\cos q$ 
(in scaled coordinates)
in addition to 
 some rapidly time varying terms. Consequently, they will adiabatically
 localize in the potential minima, that is at the center of the
resonance island.
Increasing $\gamma$ localizes successively an increasing number of states
in $\mathcal{I}_{+}.$
The switching time must be sufficiently long as compared to the 
beating period (in this case the discarded terms are still 
rapidly oscillating) but also to the inverse of the minimum energy gap 
(of the order of $\hbare^{-1}$ if $\Delta k$ is sufficiently narrow). 
In order to trap all the initially populated states, $\gamma$ has to be
sufficiently
large. For a given Bloch angle, we want to localize the~$\alpha$ first
states. 
The quantum energies of a pendulum are given by the eigenvalues of the
 Mathieu equation~\cite[chap.~20]{Abramowitz/Segun65a} (see 
figures~\ref{fig:approxpendule} and~\ref{fig:approxpendulebis}).
 For a pendulum whose hamiltonian
is
\begin{equation}
        H_{\mathrm{pend}}=\frac{p^2}{2}-\frac{\gamma}{2}\cos q\;,
\end{equation}
the phase space volume enclosed by the separatrix is
given~$16\sqrt{\gamma/2}$. 
Semiclassically, it corresponds to~$16\sqrt{\gamma/2}/(2\pi\hbare)$
states. 
This number will be of the order of $\alpha$  when $\gamma$ reaches the
value:
\begin{equation}
        \gamma_{\mathrm{adiab}}\simeq\frac{(\alpha \pi\hbare)^2}{128}\;.
\end{equation}

Figures~\ref{fig:approxpendule} and \ref{fig:approxpendulebis}
show the exact energy levels of the system together with the ones
using the pendulum approximation and the Mathieu
equation, as well as plots of selected Husimi representations for few
eigenstates.
This allows to check that :
\begin{itemize}
\item
The pendulum approximation works well in the regime of interest, say
up to $\gamma=0.1.$
\item The semiclassical estimate of the number
of  trapped states is sufficiently accurate for our purpose.
\item The Husimi representations of the trapped states
are well localized: especially, the states in
figure~\ref{fig:approxpendulebis}
have two well separated components in the $\mathcal{I}_{+}$ and
$\mathcal{I}_{-}$ islands, meaning that we are in a real case of
tunnelling.
\item During the initial increase of $\gamma$, the ``ground" state is
well
isolated (in energy) from the other ones, which means that an adiabatic
preparation is possible.
\end{itemize}

Nevertheless, for this adiabatic preparation
to be valid, we must make sure that~$\gamma$ has not reached
 a range where chaos have non negligible effects
on quantum properties. Physically, we want that chaotic layers 
at~$\gamma=\gamma_{\mathrm{adiab}}$ have a small volume compared to the
Planck 
constant.
From figure~\ref{fig:approxpendule},
 we can obtain the upper bound limit of~$\gamma$  
by estimating
when the (small) avoided
crossings become too large to be passed diabatically.
  For~$1/50<\hbare<1/5$, 
we observe that chaos have no influence 
for~$\gamma<\gamma_{\mathrm{chaos}}\simeq0.04$. 

The adiabatic preparation of all the atoms in one stable island will be 
achieved 
if~$\gamma_{\mathrm{adiab}}<\gamma_{\mathrm{chaos}}$  that is if the
atoms 
are cold 
enough  to have
\begin{equation}\label{cond:adiabatic}
                \alpha<\frac{8\sqrt{2\gamma_{\mathrm{chaos}}}}{\pi\hbare}
                \simeq\frac{0.7}{\hbare}\;.
\end{equation}
In every situation considered in the following, we will have to check  
that this condition 
is fulfilled. Next we have to reach the desired value for $\gamma$ 
($\simeq 0.18$) in 
the chaotic regime while preserving the state in the island. This will
be achieved if $\gamma$ is increased sufficiently fast so that all 
encountered avoided crossings with chaotic states are passed 
diabatically.

Hence the whole preparation of the initial atomic state proceeds by 
two steps: a first adiabatic increase of $\gamma$ at the very
beginning 
to significantly populate regular states in one island followed by 
a diabatic increase of $\gamma$ to preserve them in the chaotic regime.

\subsection{How to force tunnelling ?}

As discussed in section~\ref{subsec:tunnellingstates}, the tunnelling splitting is small only
for Bloch vector $k=0.$ As it is not presently possible to prepare
only this value of $k$ in a real experiment, it seems at first sight
that only a small fraction of atoms (close to $k=0$) may effectively
tunnel, hence considerably reducing the signal to noise ratio.
A solution is to force all atoms to go through the $k\approx0$ region.
The simplest idea is to impose a slow increase of the $k$ value, by
adding
a constant external force $F$ from outside. 
Then, the full Hamiltonian which 
governs the dynamics is:
\begin{equation}
        H'(p,q,\tau)=H(p,q,\tau)-qF\;.
\label{drift}
\end{equation}
The potential~$V$ induces a dynamical drift in the  Bloch angle:

\begin{equation}\label{eq:kpoint3}
        \frac{d}{d\tau} k(\tau)
        =\frac{1}{\hbare}F\;.
\end{equation} 
It is shown in appendix~\ref{app:banddynamics} that this 
relation, which is well-known in the time independent case 
(see for instance~\cite[chap.6]{Callaway74a}), remains valid
when~$H$ is periodic in time.

A convenient way of realizing experimentally such a constant force
is to chirp the laser frequencies, that is make all the frequencies
drift linearly in time. In an accelerated frame, the laser frequencies
appear as constant, and we are back to our model. However, in this
non inertial frame, the constant acceleration is translated in a
constant
force, hence the system is governed by equation~(\ref{drift}).
This method has been used with cold atoms, see~\cite{BenDahan+96a}.

The global result is a slow drift of the $k$ distribution. This makes
that the various $k$ classes successively come close to $k=0$ and
are thus able to tunnel. 
Whether the atom will effectively tunnel or not depends on the time
scale on which $k$ changes. If $k$ varies rapidly, the avoided
crossing at $k=0$ is crossed diabatically, i.e. the velocity
distribution will not be modified. If $k$ varies slowly, it is crossed
adiabatically (rapid adiabatic passage).  The 
Landau-Zener formula\cite{Zener32a} 
yields for the typical time scale for the crossover between
diabatic and adiabatic crossing.

\begin{figure}[!ht]
\center  
\psfig{file=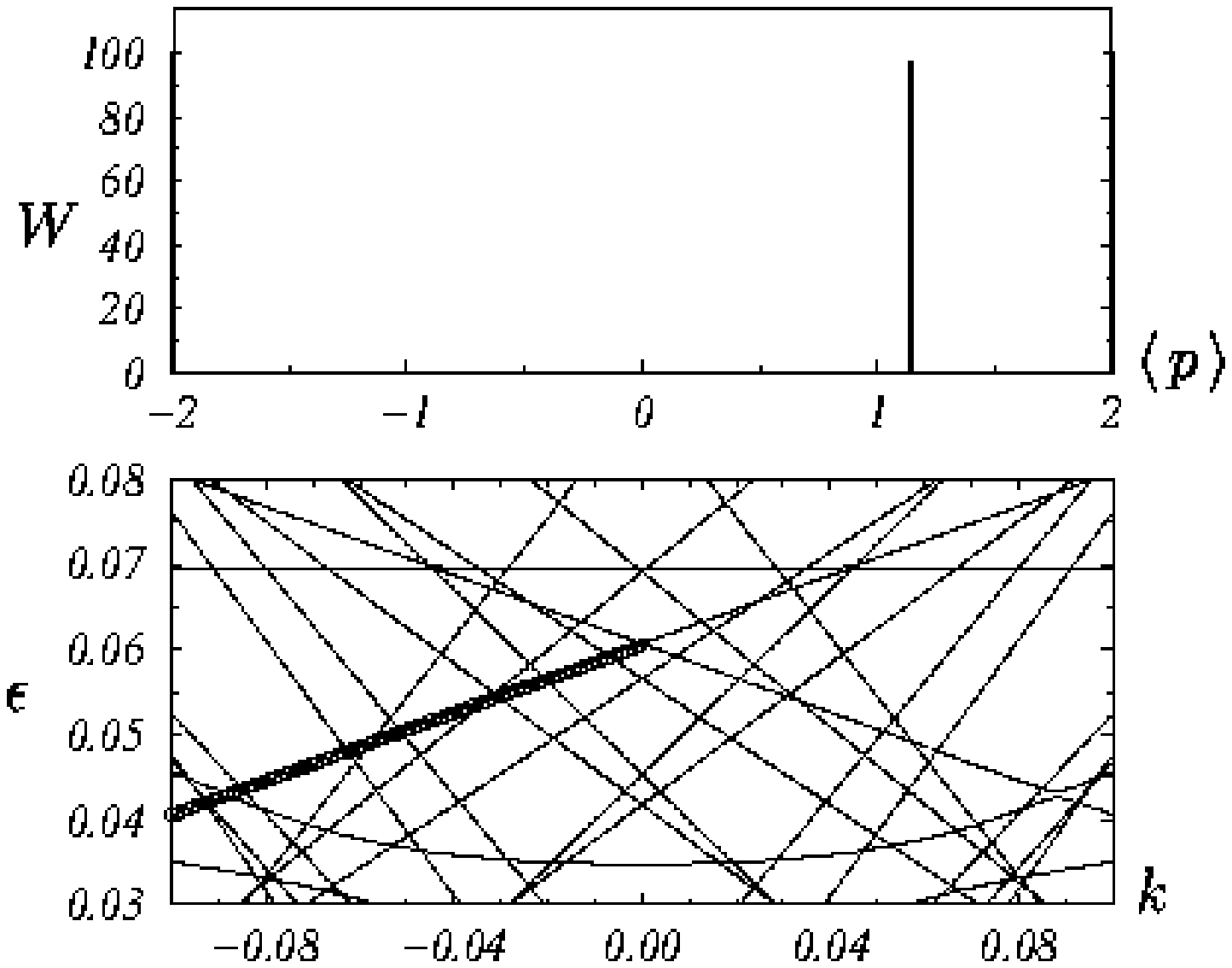}
\caption{\label{fig:suiviadiabatique} In order to measure the
tunnelling splitting at~$k=0$ (see figures~\ref{fig:bandes_un}
 and \ref{fig:bandes_deux}), the atoms are prepared into states
localised
inside the~${\mathcal{I}}_+$ resonance island. The average momentum 
distribution (top graph) is peaked
at a value slightly larger than~$+1$ in agreement with
figure~\ref{fig:3yeux}.
 The initial states occupy 1/10 of the first Brillouin zone and are
represented  
by the circles in the lower graph. 
($\hbare=0.2037,\gamma=0.18)$. 
After an adiabatic sliding of~$\Delta k\simeq0.1$, the states populated
and the 
average momentum distribution is shown in
figure~\ref{fig:suiviadiabatique2}.
}
\end{figure}

\begin{figure}[!ht]
\center  
\psfig{file=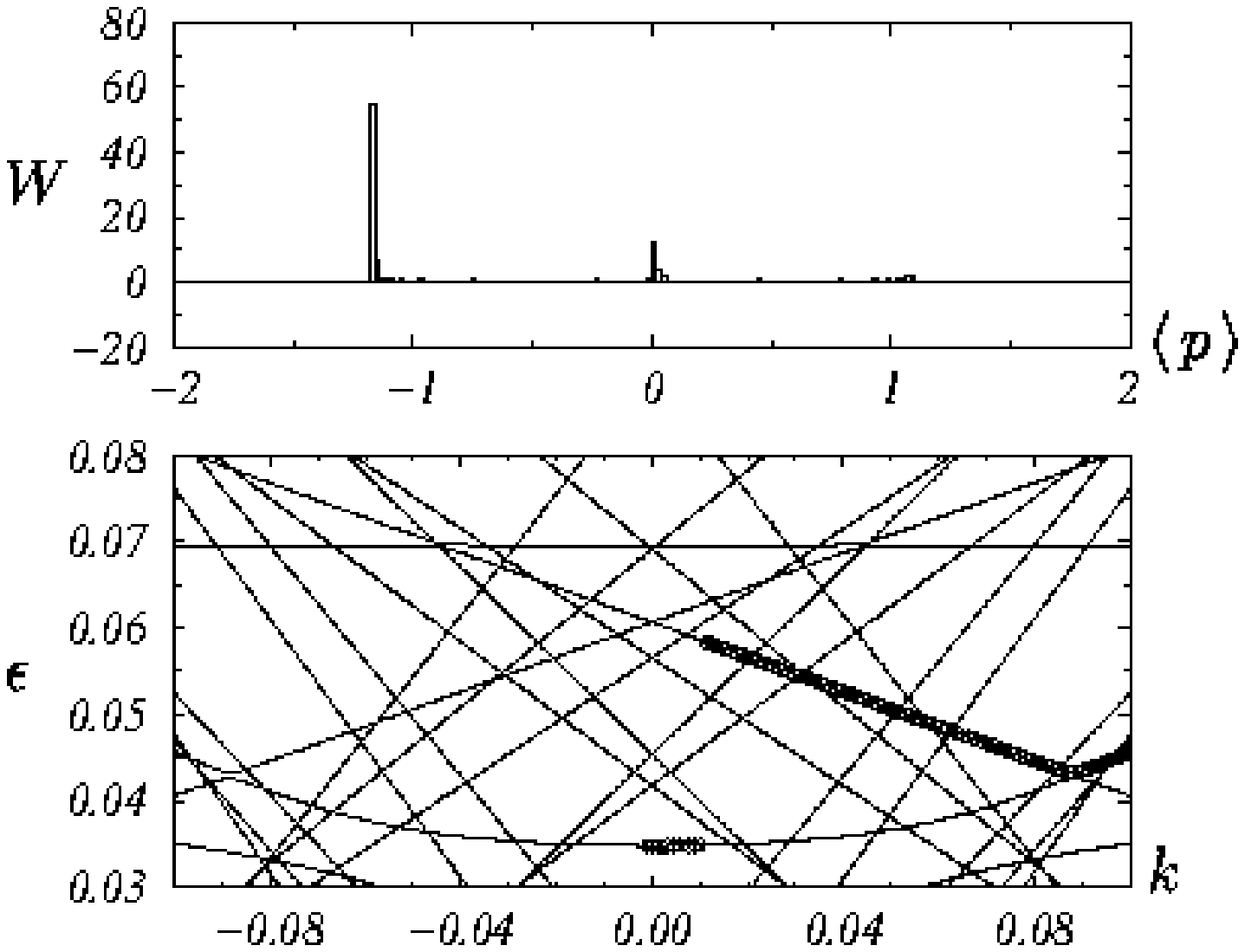}
\caption{\label{fig:suiviadiabatique2} The final states obtained
after drift of the Bloch vector by $\Delta k\simeq1/10$ 
(initial state in 
figure~\ref{fig:suiviadiabatique}). If the force
governing the motion~\eqref{eq:kpoint3} is chosen such that the 
sliding is adiabatic through the avoided crossing at~$k=0,$ the 
momentum
of the tunnelling atoms has just reversed its sign.
If the initial states cover a small enough well-centered interval 
of the Brilloin zone, only the $k=0$ avoided crossing
is important. Other avoided crossings can be seen
at~$k\simeq\pm0.09$ 
and~$\epsilon\simeq0.43$. Those are responsible for the momentum 
dispersion
of the atoms and should be avoided as much as possible.
}
\end{figure}

Figures~\ref{fig:suiviadiabatique} and~\ref{fig:suiviadiabatique2}
illustrate,
 for~$\hbare=0.2037$
 the drift of  atoms
 initially localized in~$\mathcal{I}_+$ (one energy band)
 whose distribution in~$k$ covers
1/10 of the Brillouin zone from~$k\simeq-1/10$ to $k\simeq0$ 
(sub-recoil initial velocity distribution with width of the
order of $v_{\mathrm rec}/5$)
Let us have the atoms evolving under the force~$F$ in order 
to get a global
 translation of~$1/10$ in~$k$.
After a  sliding across the splitting at~$k=0$, if 
the force~$F$ is weak to follow the energy level adiabatically,
 the average momentum of the
 atoms has reversed its sign as can be seen 
on figure~\ref{fig:suiviadiabatique2}. 
Measuring the critical value of~$F$
for this Landau-Zener like transition to occur furnishes a mean to
measure
the splitting.

Another possibility would be to modulate the external force (and thus
the
$k$ values) periodically
in time in order to induce a resonant transfer between~$\mathcal{I}_{+}$
and~$\mathcal{I}_{-}$.

In any case, the method may work only if no other avoided crossing
come into play. Numerical investigations show that there are mainly
tiny avoided crossings along the energy curve of interest. However, as
a general rule, there are usually few avoided crossings with similar
or larger size than the avoided crossing of interest. If such
an avoided crossing is also passed adiabatically, it will of course
spoil the momentum distribution. Hence, it is crucial for the initial
$k$ distribution to be sufficiently narrow to avoid this problem.
Hence, a sub-recoil velocity distribution seems necessary.

\subsection{Detection}

After the atoms have interacted with the modulated waves, one can
switch off the lasers either abruptly or adiabatically (in which case
the atoms adiabatically leave the resonance island). In both cases,
the atoms which have tunneled from $\mathcal{I}_{+}$
to~$\mathcal{I}_{-}$ will end with a momentum close to $-p_0.$
A standard time of flight technique should be enough to detect them.
A more sophisticated technique, for example based on velocity
sensitive Raman transitions~\cite{Vuletic+98a,Morinaga+99a} could also be
used with a subrecoil resolution if needed
\cite{BenDahan+96a}.

\section{Conclusion}\label{sec:conclusion}

In this paper we have proposed a simple and accessible 
experimental configuration in which the observation of chaos 
assisted tunnelling should be feasible. It consists in atoms 
propagating in the light field of two far-detuned monochromatic 
standing waves 
with slightly different frequencies.
Observing the tunnelling effect however requires sub-recoil cooling 
techniques to conveniently prepare the atomic sample together 
with a well-controlled experimental procedure (adiabatic preparation 
of the 
atomic state in one stable island followed by a drift of the 
 Bloch vector). As the tunnelling period fluctuates over 
several decades when the potential strength varies (see fig.8), 
observing these fluctuations requires a stabilization of the laser 
intensity at the level of a few percent.

\acknowledgments Ch. M., R. K. and A. M. would like to thank 
the Laboratoire Kastler Brossel
for kind hospitality. CPU time on computers has been provided by IDRIS.
Laboratoire Kastler Brossel de
l'Universit\'e Pierre
et Marie Curie et de l'Ecole Normale Sup\'erieure is
UMR 8552 du CNRS.

\appendix
\section{Derivation of the effective hamiltonian}
\label{app:effham}

To derive the effective Hamiltonian~\eqref{eq:ham_eff}, 
we basically proceed in three steps\cite{Graham+92a} :

- We first assume that any dissipation process can be safely 
ignored. Indeed for tunnelling to be observable, phase coherence 
of the atomic wave-function must be preserved during all the 
process. In our case, this means that spontaneous 
emission must be negligible. It can be shown 
\cite{CohenTannoudji90a} 
that increasing 
the laser detuning considerably decreases phase coherence 
loss. Hence the evolution will be essentially Hamiltonian 
provided each laser beam is sufficiently far-detuned from 
the atomic resonance :
\begin{equation}
|\delta_\pm|=|\omega_\pm-\omega_{\mathrm{at}}|\gg\Gamma
\end{equation}

When this holds, the total Hamiltonian operator for this 
two-level system is the sum of 
three terms: the kinetic energy operator describing the 
center-of-mass motion of atoms of mass M 
 the 
energy operator for the internal degrees of freedom and the coupling 
between internal and external degrees of freedom. 
In the dipolar approximation, 
the interaction is just~$dE(x,t)$ and does not depend on~$y$ and~$z$.
The dynamics along~$y$ and~$z$ is just trivially described by free motion and
 can be easily eliminated since the total quantum state factorizes as a
plane wave in~$y$
and~$z$. One is then left with the dynamics along~$x$ which is described
 by the hamiltonian :
\begin{equation}
        H_{\mathrm{at}}= 
        \frac{p_x^2}{2M}\big(|e\rangle\langle e|+|g\rangle\langle g|\big)
                       +\hbar\omega_{\mathrm{at}}|e\rangle\langle e|
                       -dE(x,t)\big(|e\rangle\langle g|+|g\rangle\langle e|\big)
\end{equation}
where~$p_x$ is the atomic momentum along~$x$. $|g\rangle$ and $|e\rangle$ 
are the ground-state 
and excited-state and 
$d$ is the atomic dipole strength connecting them.

- Second, we expand the total atomic state as: 
\begin{equation}
|\Psi\rangle = \psi_g(x,t)|g\rangle+\psi_e(x,t)
\exp{(-\imat \omega_Lt)}|e\rangle
\end{equation}
and we drop high frequency 
(optical) anti-resonant terms as far the amplitudes change slowly 
during an optical period.
This is known as the rotating wave approximation~\cite{Allen/Eberly87a}
and features the averaging procedure to eliminate fast variables in 
classical perturbation theory~\cite{Lichtenberg/Lieberman83a}.
This yields the coupled amplitude equations:

\begin{subequations}\label{eq:rwa}
\begin{eqnarray}
 \imat\hbar\partial_t\psi_g     &=& -\frac{\hbar^2}{2M}\partial_{xx}^2\psi_g
         -\frac{\hbar\Omega(x,t)}{2}\psi_e\;;\label{eq:rwaa}\\
 \imat\hbar\partial_t\psi_e     &=&
 -\frac{\hbar^2}{2M}\partial_{xx}^2\psi_e-\hbar\delta_L\psi_e
         -\frac{\hbar\Omega^*(x,t)}{2}\psi_g\;.
 \label{eq:rwab}
\end{eqnarray}
\end{subequations}

In these equations $\delta_L=\omega_L
-\omega_{\mathrm{at}}$ is the mean
laser detuning, the star denotes complex conjugation and  
$\Omega(x,t)$ reads:
\begin{equation}
\Omega(x,t)=\big[
           \Omega_+\exp(-i\delta\omega\,
t/2)+\Omega_-\exp(i\delta\omega\, t/2)
            \big]\cos(k_Lx)
\end{equation}
where $\Omega_\pm=d\,E_\pm/\hbar$ 
are the Rabi frequencies of each standing wave.

- As a final step, we assume
now that atoms initially prepared in their ground-state 
mostly evolve in their ground-state. This means 
that the whole atomic dynamics is solely determined by 
the ground-state amplitude $\psi_g$. For this to hold, 
adiabatic elimination of the excited-state 
amplitude~\cite{CohenTannoudji90a} must be justified. 
If the spatial partial derivatives were absent, equations 
(\ref{eq:rwa}) would just describe the Rabi oscillation phenomenon. 
It is then known that far off resonance, i.e. when the 
frequency separation of the states is much larger than any 
other frequencies, the Rabi oscillation 
is very small in amplitude. A sufficient condition is :

\begin{equation}
  |\delta_L| \gg \Omega_\pm, \delta \omega
\end{equation}

If in addition we assume that the excited-state kinetic 
energy is very small (which will be easily achieved with 
cold atoms):

\begin{equation}
        |\delta_L| \gg \left\langle\psi_e\left|
\frac{p_x^2}{2M}\right|\psi_e\right\rangle
\end{equation}
 
then adiabatic elimination of the excited-state 
amplitude just amount to neglect the spatial and temporal 
derivatives 
of $\psi_e$ in equation~\eqref{eq:rwab} which is then 
solved as $\psi_e\simeq-(\Omega^*/2\delta_L)\psi_g \ll \psi_g$. 
It is then easy to see that the 
ground-state amplitude $\psi_g$ 
obeys an effective Schr\"odinger equation with Hamiltonian:
\begin{equation}
H=\frac{p_x^2}{2M}+\frac{\hbar|\Omega(x,t)|^2}{4\delta_L}\;.
\end{equation}
Eventually, up to an irrelevant purely time-dependent term, we get
\begin{equation}
        H=\frac{p_x^2}{2M}-V_0\cos(2k_Lx)[\theta+\cos(\delta\omega\, t)]
\end{equation}
where $V_0\DEF-\hbar\Omega_+
\Omega_-/8\delta_L$ and 
$\theta\DEF(\Omega_+/2\Omega_-)
+(\Omega_-/2\Omega_+)$.

\section{Floquet-Bloch formalism}
\label{app:FBformalism}

In this appendix, we briefly recall the Floquet-Bloch formalism which is
used for a quantum problem whose Hamiltonian~$H$ is periodic 
both in space and in time. We will  denote $T$
and $Q$ the temporal and spatial periods 
respectively.

Let us first consider the time periodicity. We 
define~\cite{Floquet1883a,Shirley65a,Zeldovich67a} 
\begin{equation}\label{def:K}
        K(\opp,\opq,\tau)\DEF-\imat\hbar\frac{d}{d\tau}+
                             H(\opp,\opq,\tau)
\end{equation}
where $\opp$ and $\opq$ stand for canonical Hermitian operators whose 
commutator is~$[\opp,\opq]=-\imat\hbar$.

If $U(\tau',\tau)$ denotes the unitary evolution operator from 
$\tau$ to $\tau'$ associated with Hamiltonian $H,$
the periodicity of the dynamics
implies that~$U(\tau+T,T)=U(\tau,0)$ and
\begin{equation}
        U(\tau+T,\tau)=U(\tau+T,T)\;U(T,0)\;U(0,\tau)
                    =[U(0,\tau)]^{-1}\;U(T,0)\;U(0,\tau)\;.
\end{equation}
This shows that $U(\tau+T,\tau)$ and $U(T,0)$
  differ
 by a unitary
transformation and hence have the same spectrum (but of course
different eigenvectors), independent of $\tau$.
The eigenvalues of $U(\tau+T,\tau)$ have unit modulus
and can be written as ${\mathrm{e}}^{-\imat\epsilon_n T/\hbar}$
where $\epsilon_n$ is the so-called quasi-energy defined
modulo $2\pi\hbar/T.$ If $|\psi_n(\tau)\rangle$ denotes the
corresponding
eigenvector, we can define the Floquet state:
\begin{equation}
\label{def:fstate}              
                |{\chi_n(\tau)}\rangle\DEF
                {\mathrm{e}}^{\imat\epsilon_n \tau/\hbar}|
                {\psi_n(\tau)}\rangle
\end{equation}
which is by construction periodic with period $T$. 

Inserting the definition of the Floquet state 
in the time-dependent Schr\"odinger equation, we immediately obtain:
\begin{equation}
                K(\opp,\opq,\tau)\,|{\chi_n(\tau)}\rangle
                =\epsilon_n\,|{\chi_n(\tau)}\rangle\;.
\end{equation}
which means that the quasi-energy spectrum is obtained by diagonalizing
the Floquet Hamiltonian in the space of time-periodic functions.

The second step consists in making use of the invariance of $H$ under 
spatial translations with period $Q$. 
The unitary translation operator~$\widehat{T}_Q\DEF
{\mathrm{e}}^{-\imat \opp Q/\hbar}$
commutes with~$K$. We can then use the spatial counterpart of the
Floquet 
theorem, namely the Bloch theorem \cite{Ashcroft/Mermin76a}, 
and label the eigenstates of $K$ with 
the Bloch number~$k\in[-\pi/Q,\pi/Q[$ (the first Brillouin zone),
which means diagonalising $K$ in each subspace with fixed $k.$
If one defines the Floquet-Bloch states as:

\begin{equation}
        |{u_{n,k}(\tau)}\rangle={\mathrm{e}}^{-\imat k
\opq}|{\chi_{n,k}(\tau)}\rangle
        ={\mathrm{e}}^{\imat\epsilon_{n}(k) \tau/\hbar}
                             {\mathrm{e}}^{-\imat k \opq}
|{\psi_{n,k}(\tau)}\rangle\;,
\end{equation}
where~$\{{\psi_{n,k}(\tau)}\rangle\}$ forms a complete orthogonal eigenbasis, it is easy to show that they can be obtained by diagonalizing the
Floquet-Bloch Hamiltonian:
\begin{equation}
\tilde{K}(\opp,\opq,\tau, k) = K(\opp+\hbar k,\opq,\tau)
\end{equation}
on the subspace of time and space periodic functions.
In our specific case, the
Floquet-Bloch Hamiltonian reads:
\begin{equation}\label{eq:tildeK}
\tilde{K}(\opp,\opq,\tau, k) = \frac{(\opp+\hbar k)^2}{2} -\gamma
\ (\theta + \cos\tau )\  \cos \opq
-\imat\hbar\frac{d}{d\tau}\;.
\end{equation}

The spatial periodicity of the Floquet-Bloch states leads to a discrete
set of
dispersion relations~$\epsilon_n(k)$ . 
For fixed~$n$, the set of all 
quasi-energies~$\epsilon_n(k)$ for
$k$~in the first Brillouin zone~$[-\pi/Q,\pi/Q[$ is called 
the~$n\mathrm{th}$ band of the system.

Let us now obtain the velocity
theorem~\eqref{eq:averagev}. Using the above relations, we have
$\langle\psi_{\epsilon,k}(\tau)|\,\opp\,|{\psi_{\epsilon,k}(\tau)}\rangle=
\langle u_{\epsilon,k}(\tau)|\,(\opp+\hbar
k)\,|{u_{\epsilon,k}(\tau)}\rangle= \langle
u_{\epsilon,k}(\tau)|\big(\hbar^{-1}\partial \tilde{K}/\partial
k\big)|{u_{\epsilon,k}(\tau)}\rangle $. The derivation with respect
to~$k$
of the relation~$\langle
u_{\epsilon,k}(\tau)|\,\tilde{K}\,|{u_{\epsilon,k}(\tau)}\rangle=\epsilon(k)$
leads to
 \begin{multline}\label{eq:psippsi}
\langle\psi_{\epsilon,k}(\tau)|\,\opp\,|{\psi_{\epsilon,k}(\tau)}\rangle\\=
         \frac{1}{\hbar}\frac{\partial \epsilon}{\partial k}
         +\imat\hbar\frac{d}{d\tau}\left(
                                \langle u_{\epsilon,k}(\tau)|
                                \frac{\partial}{\partial k}
                                |{u_{\epsilon,k}(\tau)}\rangle
                                \right)
         -\left(
                \frac{1}{\hbar}\frac{\partial}{\partial k}\langle
u_{\epsilon,k}(\tau)|
          \right)
          \tilde{K}\,|{u_{\epsilon,k}(\tau)}\rangle
         -\left[\left(
                \frac{1}{\hbar}\frac{\partial}{\partial k}\langle
u_{\epsilon,k}(\tau)|
                 \right)
                  \tilde{K}\,|{u_{\epsilon,k}(\tau)}\rangle
          \right]^*\;. 
\end{multline} 
The two last terms of the the right hand
side
are opposite since the normalization of the $u$'s leads to
$\partial\langle u_{\epsilon,k}|u_{\epsilon,k}\rangle /\partial k=0$.
Moreover, after time-averaging~\eqref{eq:psippsi} over~$T$, the total
$\tau$-derivative vanishes since the $u$'s are precisely~$T$-periodic
while the time-independent $k$-derivative of the quasi-energy remains
unchanged. Eventually,
 \begin{equation}
        \frac{1}{T}\int_0^T  
\langle\psi_{\epsilon,k}(\tau)|\,\opp\,|{\psi_{\epsilon,k}(\tau)}\rangle
        \,d\tau=\frac{1}{\hbar}\frac{\partial \epsilon}{\partial k}\;.
\end{equation}

\section{Bloch angle dynamics}
\label{app:banddynamics}
In this appendix we derive equation~(\ref{eq:kpoint3}) 
which is valid for an arbitrary 
strength of the constant force~$F$ provided that the potential~$V=-Fq$ remains
strictly linear in~$q$.
Let us choose a state~$|\psi(\tau)\rangle$ evolving under~$H'=H+V$ such 
that it coincides with a Floquet-Bloch state at~$\tau=0$:

\begin{equation}
	\imat\hbare\frac{d\,|{\psi(\tau)}\rangle}{d\tau}
        =H'(\opp,\opq,\tau)\,|{\psi(\tau)}\rangle
\end{equation}
and
\begin{equation}
	|\psi(\tau=0)\rangle=|\psi_{n,k}(\tau=0)\rangle\;.
\end{equation}
In the interaction picture we have immediately
\begin{equation}\label{eq:intpicture1}
	\imat\hbare\frac{d\,|{\psi^I(\tau)}\rangle}{d\tau}
        =-F U^\dagger(\tau,0)\opq U(\tau,0)\,|{\psi^I(\tau)}\rangle
\end{equation}
where~$|{\psi^I(\tau)}\rangle\DEF U^\dagger(\tau,0)|\psi(\tau)\rangle$
and where~$U$ denotes the evolution operator under~$H$.
Let~$|\phi(\tau)\rangle$ be the ket defined by
\begin{equation}\label{eq:intpicture2}
	|\phi(\tau)\rangle\DEF{\mathrm{e}}^{-\imat F\opq \tau/\hbare}|{\psi^I(\tau)}\rangle\;.
\end{equation}
It is straightforward to obtain its evolution :
\begin{equation}
	\imat\hbare\frac{d\,|{\phi(\tau)}\rangle}{d\tau}
        =G(\tau)\,|{\phi(\tau)}\rangle
\end{equation}
where
\begin{equation}
	G(\tau)\DEF F\left(\opq-{\mathrm{e}}^{-\imat\tau F\opq/\hbare}
                                \,U^\dagger(\tau,0)\,\opq\, U(\tau,0)
                     		{\mathrm{e}}^{\imat\tau F\opq/\hbare}
                     \right)\;.
\end{equation}
Since~$U^\dagger(\tau,0)$ commutes with the translation 
operator~$\widehat{T}_Q$, it can be checked that~$[G(\tau),\widehat{T}_Q]=0$.
The evolution of~$|{\phi(\tau)}\rangle$ under~$G$ will therefore  preserve its initial 
 quantum number~$k$ :
\begin{equation}
	\widehat{T}_Q|{\phi(\tau)}\rangle={\mathrm{e}}^{-\imat k Q}|{\phi(\tau)}\rangle
\end{equation}
for all~$\tau$. Thus, making use of~\eqref{eq:intpicture1} and~\eqref{eq:intpicture2}, we have
\begin{equation}
	\widehat{T}_Q|{\psi(\tau)}\rangle
        ={\mathrm{e}}^{-\imat (k+F\tau/\hbare) Q}|{\psi(\tau)}\rangle
\end{equation}
which shows that~$|{\psi(\tau)}\rangle$ is actually a Bloch wave with
 a Bloch angle given by~$k(\tau)=k(0)+F\tau/\hbare$ even if it spreads
 among the quasi-energy bands.

\newpage \ 


\end{document}